\documentclass[12pt,pre]{revtex4}

\usepackage{amsmath}
\usepackage{graphicx}
\usepackage{subfigure}
\usepackage{verbatim}

\newcommand{\lb}{{<}}
\newcommand{\rb}{{>}}
\newcommand{\BK}{Burridge-Knopoff}
\newcommand{\bk}{Burridge-Knopoff}
\newcommand{\OFC}{Olami-Feder-Christensen}
\newcommand{\RJB}{Rundle-Jackson-Brown}
\newcommand{\ca}{cellular automata}

\newcommand{\nn}{nearest-neighbor}
\newcommand{\lr}{long-range}
\newcommand{\mf}{mean-field}
\newcommand{\nmf}{near-mean-field}
\newcommand{\st}{stress transfer}

\newcommand{\bb}{x}
\newcommand{\bs}{\bb_{\rm s}} % number of failed sites
\newcommand{\bm}{\bb_{\rm m}} % moment
\newcommand{\df}{\lb\Delta f\rb}
\newcommand{\du}{\lb\Delta u\rb}

\begin{document}

\title{Near mean-field behavior in the generalized \BK\ earthquake model with variable range \st}
\author{Junchao Xia}
\altaffiliation[Present address: ]{Department of Chemistry, The University of Iowa,
Iowa City, IA 52240}
\affiliation{Department of Physics, Clark University, Worcester, MA 01610}

\author{Harvey Gould}
\affiliation{Department of Physics, Clark University, Worcester, MA 01610}

\author{W. Klein}

\affiliation{Department of Physics and Center for Computational
Science, Boston University, Boston, MA 02215}
\author{J. B. Rundle}
\affiliation{Department of Physics and Center for Computational
Science and Engineering, University of California, Davis, CA 95616}

\begin{abstract}

Simple models of earthquake faults are important for understanding the
mechanisms for their observed behavior in nature, such as Gutenberg-Richter scaling. Because of the importance of 
long-range interactions in an elastic medium, we generalize the \bk\ slider-block model to include variable range stress transfer. 
We find that the \bk\ model with \lr\ \st\ exhibits qualitatively different
behavior than the corresponding \lr\ cellular automata models and the usual \BK\ model with nearest-neighbor
stress transfer, depending on how quickly the friction force weakens with increasing
velocity. Extensive simulations of 
quasiperiodic characteristic events,
mode-switching phenomena, ergodicity, and waiting-time distributions are also discussed.
Our results are consistent with the existence of a \mf\ critical point and have important implications for our understanding of earthquakes 
and other driven dissipative systems.

\end{abstract}

\maketitle

\section{Introduction}
\label{sect:introduction}

Earthquake faults are important examples of driven dissipative
systems~\cite{GeoComp00}. Models of fault systems are important for
understanding Gutenberg-Richter (power law) behavior, the occurrence of characteristic
events, and the relation between small and large earthquakes~\cite{GeoComp00,
ScholzBook02, BowmanJGR98,GRLaw, FisherPRL97}. 
Understanding driven dissipative systems is important, for example, in the context of avalanches~\cite{BTWSOC}, 
neural networks~\cite{HertzPRL95}, 
depinning transitions in charge density waves and superconductors~\cite{DFisherPR98}, 
magnetized domains in ferromagnets~\cite{UrbachPRL95}, domain rearrangements in flowing
foams~\cite{GopalPRL95}, and granular materials under shear
stress~\cite{OhernPRL04}.

A relatively simple dynamical model that contains much of the essential physics of earthquake faults
is the spring-block model proposed by Burridge and Knopoff~\cite{BKModel67}.
This model consists of blocks connected by linear springs 
to their nearest neighbors with spring constant $k_c$. 
The blocks are also connected to a loader plate by linear springs
with spring constant
$k_L$, and rest on a surface with a nonlinear velocity-weakening stick-slip
friction force which depends on a parameter $\alpha$ that controls how
quickly the friction force decreases as the velocity is increased. The
model was studied numerically in one dimension in Ref.~\cite{BKModel67} and
more recently in Refs.~\cite{CarlsonPRL89,CarlsonPRA89,CarlsonPRA91,LangerPRL91,CarlsonRMP94,Nagelgroup,clancy,MoriBK, MoriBK2}. 

An earthquake event is defined as a cluster of blocks that move (slip) due
to the initial slip of a single block. In addition to the amount of energy released in an earthquake event, a
quantity of interest is the moment $M$, which is defined as
$\sum_j \Delta x_j$, where $\Delta x_j$ is the net displacement of block $j$ during an event and the sum is over all the blocks in the event. Carlson
and Langer simulated the one-dimensional \BK\ model for $N=100$
and $N=1000$ blocks. The main
result of their simulations~\cite{CarlsonPRL89,CarlsonPRA89,CarlsonPRA91,LangerPRL91,CarlsonRMP94,clancy} 
is that for $\alpha \gtrsim 2$ the moment probability distribution $P(M)$ scales as
$M^{-\bb}$ for small localized events with an exponent $\bb \approx 2$~\cite{exponent}. There
also is a peak in $P(M)$ for large events indicating a significant presence
of characteristic (non-power law) events.

Because simulations of the \BK\ model require solving Newton's equations of
motion and are time consuming, several cellular automata (CA)
models have been proposed that neglect the inertia of the blocks
and simplify the effect of the friction force by assuming that the motion is overdamped.
These cellular automata include those due to Rundle, Jackson, and Brown~\cite{RundleJSP91} and
Olami, Feder, and Christensen~\cite{OFCModelPRL92}. 
In these models $P(s)$, the distribution of the number of blocks in an event,
does not exhibit power law scaling~\cite{GrassbergerPRE94} for \nn\ \st\ if periodic boundary conditions are used~\cite{multiscaling}.
A generalization of these CA models to include more realistic
\lr\ stress transfer~\cite{LRNature} yields considerable differences with the \nn\ CA
models~\cite{RundleJSP93} and with the original
\BK\ model. In particular, for \lr\ \st\ $P(s)$ exhibits Gutenberg-Richter 
scaling consistent with the system being near a \mf\ (spinodal) critical
point~\cite{KleinGeoComp00,RundlePRL95,KleinPRL97,FergusonPRE99}. 
In addition, the \lr\ CA models can be described by a 
Langevin equation~\cite{KleinErgodicity,KleinPRL97,KleinGeoComp00} and small and medium size events can be interpreted as
fluctuations about a free energy minimum~\cite{KleinGeoComp00,RundlePRL95,bigklein}.
Large events drive the system out of equilibrium from which the system
decays back to an equilibrium state~\cite{KleinGeoComp00}.

The CA and \BK\ models lack several elements that would make
them more realistic. In particular, the \lr\ CA models do not include inertia and more realistic 
friction laws, and the \BK\ model does not include \lr\ stress transfer. Both types of models 
do not include elastic (seismic) wave radiation (phonons) because there is no medium in which seismic waves 
can propagate. However, the lack of seismic waves is a reasonable approximation, because seismic waves 
carry little energy in real faults~\cite{seismic}.

In this paper we discuss our extensive simulations of a generalized \BK\ model 
with \lr\ stress transfer between the blocks~\cite{JunPRL}.
For various values of the dynamic friction parameter $\alpha$ and the range of stress transfer $R$,
we observed phenomena similar to real fault network systems, including
Gutenberg-Richter scaling, quasiperiodic characteristic events,
and mode-switching. Our primary results are that the behavior of the \lr\ \BK\ model differs significantly from the short-range \BK\ model, the behavior of the \lr\ \BK\ and CA models is similar only for small $\alpha$, and the nature of the friction force is important and strongly affects the behavior of the \BK\ model. In particular, we find numerical evidence for two types of scaling behavior: a \mf\ spinodal critical point similar to that found in the \lr\ CA models~\cite{KleinGeoComp00, RundlePRL95, KleinPRL97, FergusonPRE99} 
for $R\gg 1$ and $\alpha \lesssim 1$~\cite{JunPRL} and the scaling behavior found in
Refs.~\cite{CarlsonPRA89,CarlsonPRA91, clancy} for $\alpha \gtrsim 2$ and all values of $R$ studied in the range $1 \leq R \leq 500$.

\section{\BK\ Model}\label{sec:BKModel}
The original \BK\ model in
one dimension is governed by the equation of
motion~\cite{BKModel67,CarlsonPRA89,CarlsonPRA91}
\begin{equation}
\label{eq:BK1D1}
m \frac{d^2 x_j}{dt^2} = k_c (x_{j+1} - 2x_j + x_{j-1}) -
k_Lx_j - F(v + {\dot x_j}),
\end{equation}
where $x_j$ is the displacement of block $j$ from its equilibrium position,
$v$ is the speed of the substrate, which moves to the left,
$F(\dot x) = F_0\phi(\dot x/\tilde v)$ is a velocity-dependent friction
force, $\tilde v$ is a characteristic velocity, and $m$ is the mass of a block. 
The loader plate is fixed.

\begin{figure}[t] % fig1
\begin{center}
\includegraphics[scale=0.60]{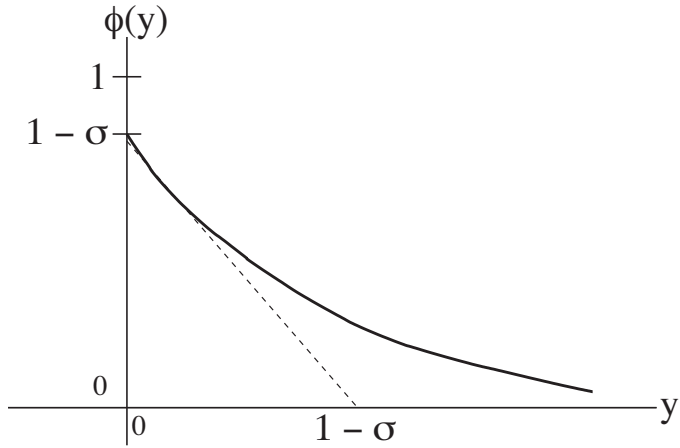}
\vspace{-1pc}
\caption{\label{fig:friction} The form of the velocity-weakening friction force $\phi(y)$.
The friction force decays monotonically to zero from the initial value $\phi(0^+)= 1 - \sigma$ with the initial slope $-1$.}
\end{center}
\vspace{-1pc}
\end{figure}

As in Ref.~\cite{CarlsonPRA89} we introduce the scaled variables $\tau = \omega_{p}t$, 
$\omega^{2}_{p} = k_L/m$, and $u_j = (k_L/F_0)x_j$, and rewrite 
Eq.~(\ref{eq:BK1D1}) in dimensionless form as
\begin{equation}
\label{eq:BK1D2}
\ddot u_j = \ell^{2}(u_{j + 1} -2u_j + u_{j - 1}) - u_j -
\phi(2\alpha \nu + 2\alpha{\dot u}_j),
\end{equation}
with $2\alpha = \omega_{p}F_0/k_L\tilde v$, $\ell^{2} = k_c/k_L$, and $\nu = v k_L/(\omega_{p}F_0)$; 
a dot denotes differentiation with respect to $\tau$. 
The form of the friction force is plotted in Fig.~\ref{fig:friction} and is given by~\cite{CarlsonPRA91}
\begin{equation}
\label{Friction}
\phi(y)=
\begin{cases}
(-\infty, 1], & \text{$y = 0$}\\
\displaystyle \frac{1 - \sigma}{\displaystyle 1 + \frac{y}{1 - \sigma}}, & \text{$y > 0$}.
\end{cases}
\end{equation}
Note that $\phi(y)$ decays monotonically to zero from $\phi(0^+)=1-\sigma$ and prohibits slip in the same direction as the motion of the substrate~\cite{fixedloadingplate}.

The four dimensionless parameters, $\ell$, $\alpha$, $\nu$, and $\sigma$ govern the behavior of the system. The parameter $\alpha$ appears in the argument of $\phi$ in Eq.~\eqref{eq:BK1D2} and determines how quickly
the dynamic friction force decreases with increasing velocity;
$\alpha = 0$ means that the dynamic friction force is equal to the constant $1-\sigma$. Larger $\alpha$ means that the friction
 force decreases more rapidly with velocity and the motion is less damped; 
$\alpha \to \infty$ implies that the friction force drops to zero immediately for positive velocities.

We generalize the \BK\ model by assuming that a block is connected to 
$R$ neighbors (in each direction) with the rescaled spring constant $k_c/R$~\cite{kac};
$R=1$ corresponds to the usual \BK\ model. We used the second- and
fourth-order Runge-Kutta algorithms~\cite{GouldBook06,NumRecipes2nd} with the time step $\Delta t = 0.001$ to
solve Eq.~\eqref{eq:BK1D2} generalized to arbitrary $R$. Both algorithms and other
fourth-order algorithms~\cite{ShampineRK4} give similar results.

\begin{comment}
discuss their results for $N=1000$ ]
JX: The main results for $N=1000$ of Calson are the same as $N=100$. But she
used zero velocity limit and obtained $P(s)$ and other quantities. 
\end{comment}

\section{Implementation of the \BK\ Model}
\label{sec:Implementation}

Because the velocity of a block is a continuous variable, we need to introduce
a criterion for when a ``stuck'' block begins to move and when a moving block
becomes stuck so that we can define the beginning and end of an event. We define 
a block to be stuck if its velocity is less than a parameter $v_0$. 
In addition, the stress on the block, defined to be the force due
to all the springs coupled to it including the loader plate spring, must be 
smaller than the maximum static friction force $F_0$ (taken to be unity in 
dimensionless units). If a block is stuck, we choose the value of the static friction 
force to be such that it cancels the stress. At the next time step, a stuck block will
remain stuck if the stress on it is still smaller than $F_0$. A moving block will become
stuck at the next time step if its speed is less than $v_0$ and decreasing and 
if the stress on it is less than $F_0$. In our simulations we take $v_0 =
10^{-5}$, which yields reasonable results.

An earthquake event
begins with the slip of a block and ends
when all blocks become stuck. A block is said to ``fail'' when it begins to move after being stuck.
A moving block can become stuck and then move (slip) again
during an event.

We initially set $\dot u_j = 0$ for all
$j$ and assign small random displacements to all the blocks; hence all
blocks are initially stuck. We compute the force on all the blocks and
update
$\dot u_j$ and
$u_j$ for all $j$ using the generalization of Eq.~\eqref{eq:BK1D2} for
arbitrary
$R$. We continue these updates until all blocks become stuck again. We
then move the substrate (the loader plate is fixed) until the stress on one
block exceeds
$F_0$. This stress loading mechanism is known as the zero velocity
limit~\cite{CarlsonPRA91,FergusonPRE99, zerovelocitylimit} and is equivalent to setting
$\nu=0$ in Eq.~\eqref{eq:BK1D2}. The zero velocity limit is realistic
because the dynamics of earthquake faults can be divided into continuous
loading on a long time scale and relaxation with release of stress and
energy on a much shorter time scale. The zero velocity limit ensures that there is only one event per substrate
update and saves considerable simulation time.
The relaxation process occurs when the 
motion or failure of the initiator induces other blocks to slip. If there are ``moving'' blocks,
the event is still alive; otherwise, we reload the system to induce a new event.

The results in this paper are for $\sigma=0.01$, $\ell = 10$, 
$v_0=10^{-5}$, $N=5000$, $10^6$ events, and various values of $R$ and
$\alpha$. Most previous work has been for $R=1$
with $N=100$~\cite{CarlsonPRA89}, $N=1000$~\cite{CarlsonPRA91}, $N = 800$~\cite{MoriBK, MoriBK2},
and the same values of $\sigma$ and $\ell$. Open boundary conditions are used as in previous work~\cite{CarlsonPRA89,CarlsonPRA91,MoriBK, MoriBK2}.

\begin{figure}[t] % fig2
\begin{center}
\includegraphics[scale=0.95]{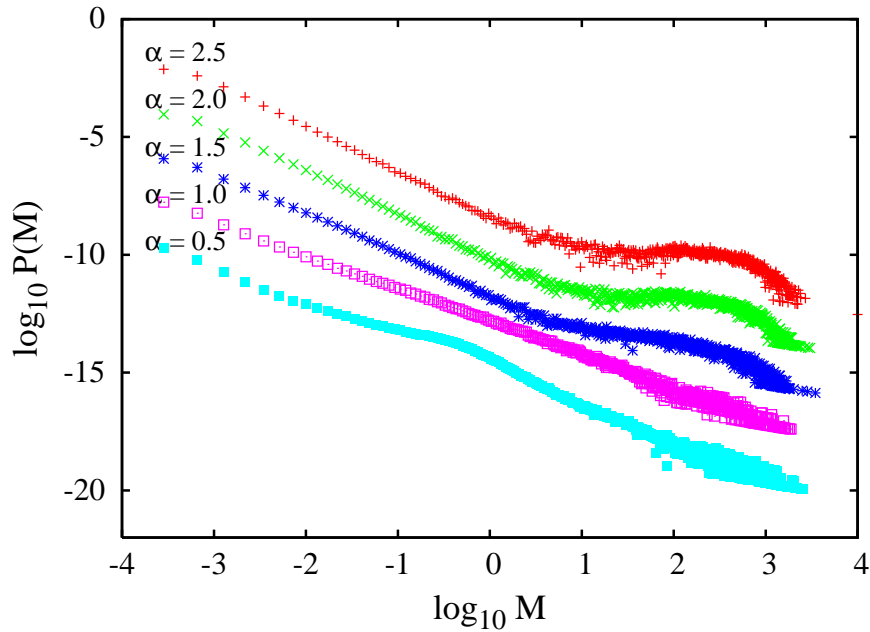}
\vspace{-1pc}
\caption{\label{fig:PM}(Color online) Log-log plot of $P(M)$ versus $M$ for different values of $\alpha$ for $R = 1$.
Scaling behavior can be found for small events for $\alpha \gtrsim 1$. The exponent $x$ defined by $P(M) \sim M^{-x}$
decreases from $x \approx 2$ as $\alpha$ is decreased from $2.5$ to $1$.
For $\alpha \lesssim 1$, no scaling behavior is observed. A nonuniform bin size was used here and in Figs.~\ref{fig3} and ~\ref{fig:PS}.
For clarity each distribution is shifted vertically by 2 units.}
\end{center}
\vspace{-1pc}
\end{figure}

\section{The Size Distribution of Earthquake Events}
\label{subsect:SizeDistr}

In Refs.~\cite{CarlsonPRA89,CarlsonPRA91} the moment distribution
$P(M)$ was found to exhibit small localized
events and larger delocalized events (for $M >2\ell/\alpha$). The localized events show
power-law behavior with a slope $\bb \approx 2$; the delocalized
or characteristic earthquakes correspond to a pronounced peak in $P(M)$.
Carlson and Langer~\cite{CarlsonPRA89,CarlsonPRA91} considered $\ell = 6$,
8, 10, and 12 and $\alpha=2.5$, 3, 4, and 5, and found the same general behavior for $P(M)$.
They also found that the effective value of the exponent $\bb$ decreases as
$\alpha \to 1$ and the power law behavior becomes less well defined.

Figure~\ref{fig:PM} shows our results for the moment distribution $P(M)$ for $R=1$ and values of $\alpha$ 
in the range $0.5 \leq \alpha \leq 2.5$. We find that for $\alpha \gtrsim 1$, $P(M)$ exhibits power law 
behavior with slope $\bb \approx 2$ for small localized events in the range $10^{-4} \lesssim M \lesssim 10^{0}$ 
and a non-power law distribution of characteristic events. 
For $\alpha \lesssim 1$, no power law behavior is found~\cite{creep}. These results are 
consistent with the results in Refs.~\cite{CarlsonPRA89, CarlsonPRA91, clancy}.

\begin{figure}[b] % fig3
\begin{center}
\includegraphics[scale=0.90]{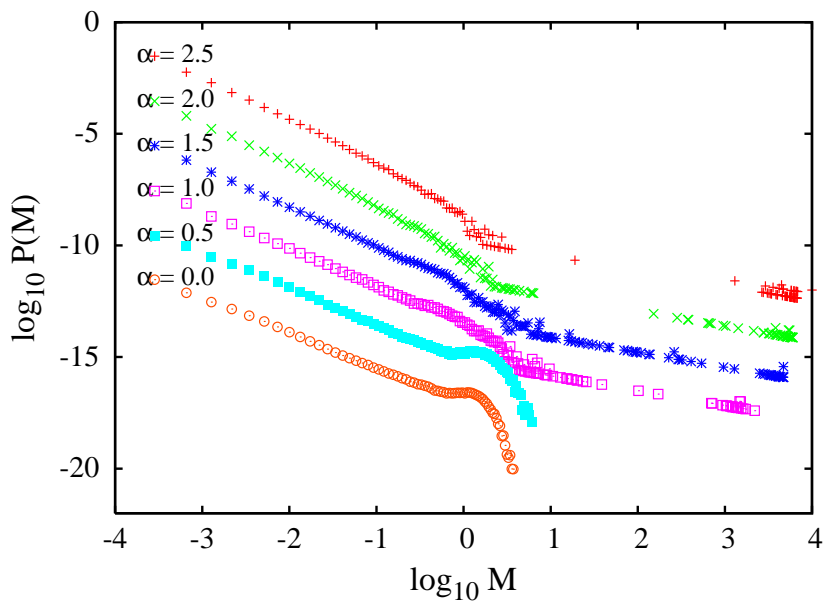}
\caption{\label{fig3}(Color online) Log-log plot of $P(M)$ versus $M$ for $R = 500$ and the same values of $\alpha$ 
as in Fig.~\ref{fig:PM}. For $\alpha \lesssim 1$ the power law exponent converges to the \mf\ value of $x=1.5$ 
as $\alpha \to 0$. 
For $\alpha \gtrsim 1$, the scaling exponent becomes close to $2$ as $\alpha$ is increased. 
The scaling range becomes smaller as $\alpha$ increases. For clarity each distribution is 
shifted vertically by 2 units. (The apparent linear behavior for $M \gtrsim 1$ is an artifact due to limited statistics and the use of a nonlinear bin width.)}
\end{center}
\vspace{-1pc}
\end{figure}

Figure~\ref{fig3} shows $P(M)$ for $R=500$ and the same values of $\alpha$ as in Fig.~\ref{fig:PM}.
For $\alpha \gtrsim 1$ we see that the increased interaction range
$R$ does not change the value of the exponent $\bb$, but the scaling range
becomes smaller, and the distribution of characteristic events ($M \gtrsim 1$)
is concentrated in a narrower range close to the system size. We conclude that although $P(M)$ exhibits power law behavior for $\alpha = 2.5$,
the slope $\bb \approx 2$ differs from the \mf\ value of $\bb=3/2$ found in the \lr\ 
CA models~\cite{KleinPRL97,FergusonPRE99}.

In contrast, for $R \gtrsim 100$ and $\alpha \lesssim 1$ the scaling exponent approaches 
$x \approx 1.5$ as $\alpha \to 0$.
This value of $x$ is close to the \mf\ value of 3/2 found in the \lr\ CA models. Increasing $R$ 
increases the range of the power law behavior and decreases the number of characteristic events.
For $\alpha = 0$ the \mf\ behavior of $P(M)$ becomes even better defined. We conclude that the scaling
exponent of $P(M)$ converges to the \mf\ value of 3/2 for sufficiently small values of $\alpha$ 
and sufficiently large values of $R$. 

\begin{figure}[t] % fig4
\begin{center}
\subfigure[\label{fig:PS.a}]{
\includegraphics[scale=0.63]{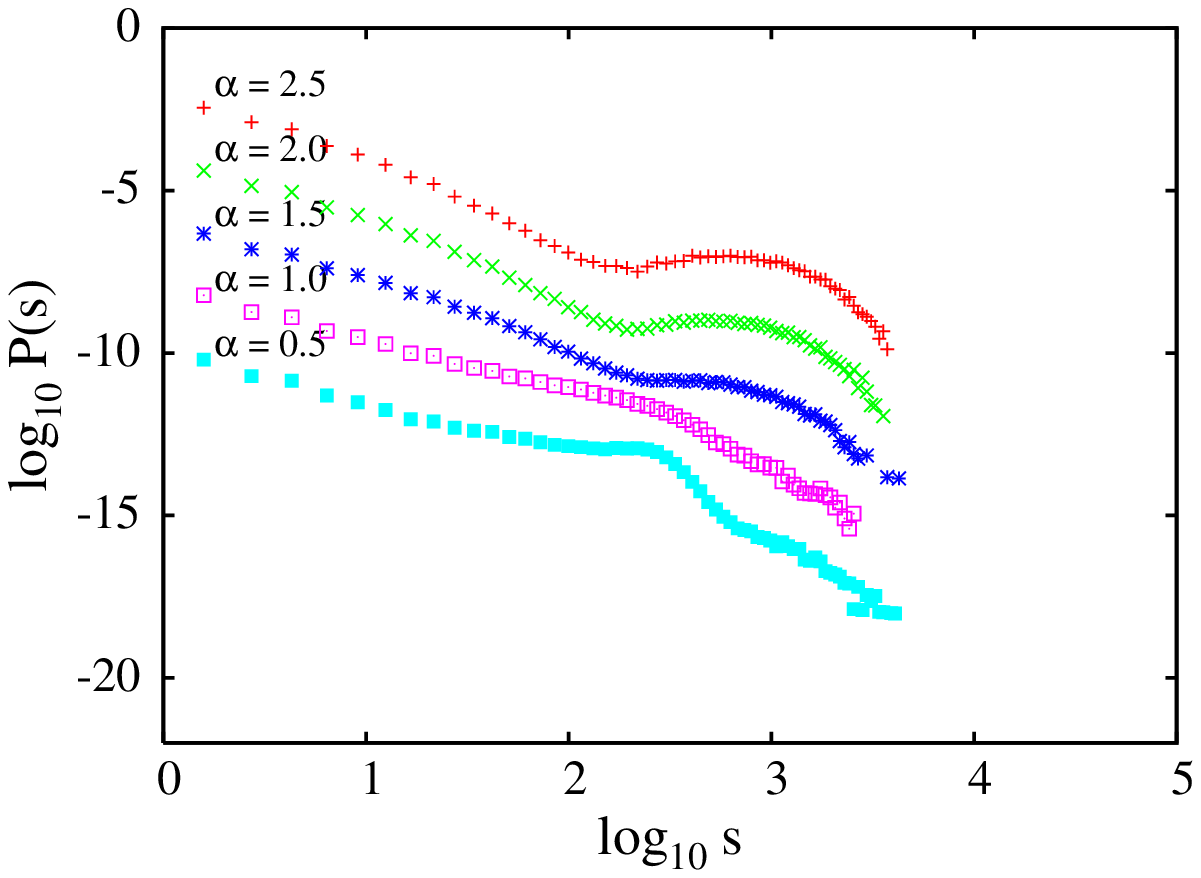}}
\subfigure[\label{fig:PS.b}]{
\includegraphics[scale=0.63]{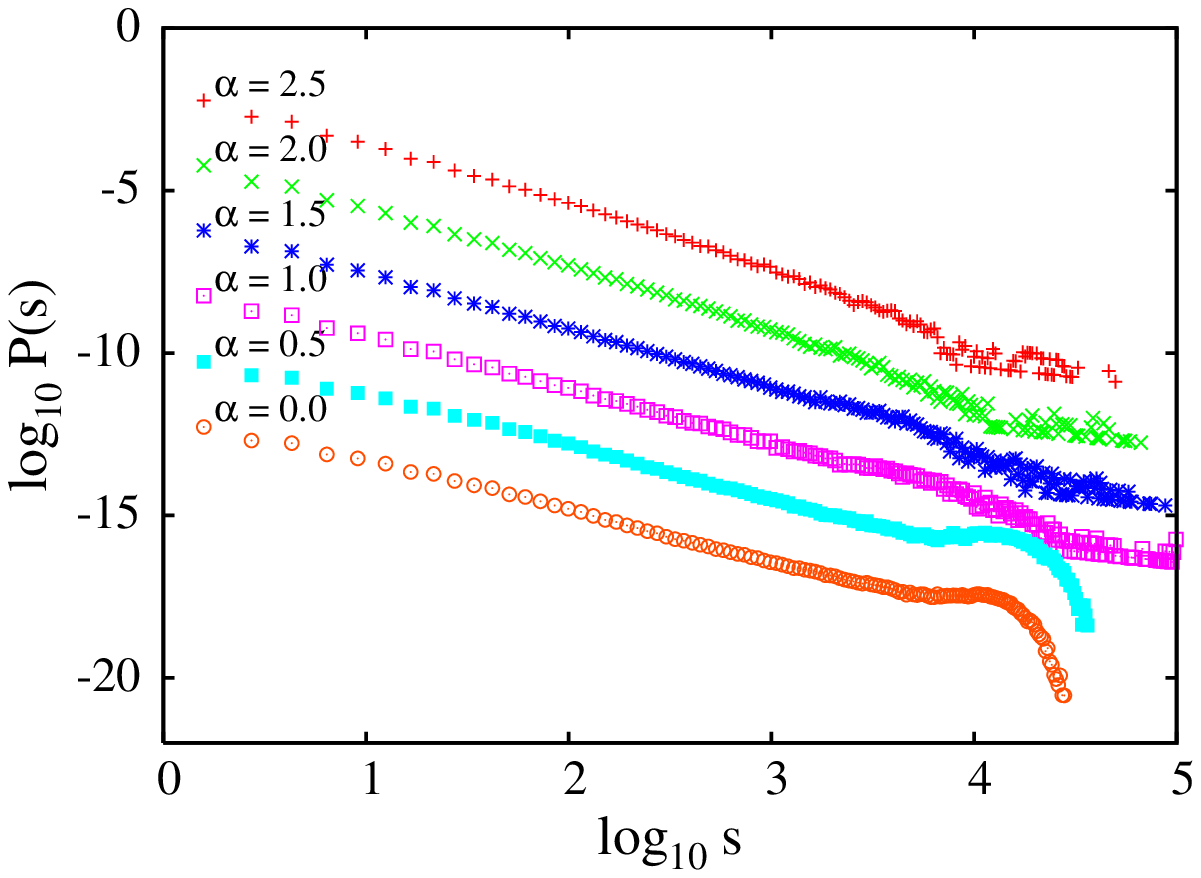}}
\vspace{-1pc}
\caption{\label{fig:PS}(Color online) The distribution of events with $s$ blocks, $P(s)$. 
(a) Log-log plot of $P(s)$ versus $s$ for $R = 1$ and various values of $\alpha$. 
In contrast to the behavior of $P(M)$, $P(s)$ does not exhibit power law behavior for
$R=1$ for all values of $\alpha$ studied. (b) For $R = 500$, $P(s) \sim s^{-\bb}$ 
with $\bb \approx 2$ for $\alpha \approx 2.5$ and $\bb \approx 1.5$ for $\alpha \lesssim 0.5$. Note that $s$ can exceed $N$ because a block can slip and become stuck 
and then slip again during an event. For clarity each distribution is shifted vertically by 2 units.}
\end{center}
\end{figure}

\begin{table}[h]
\begin{center}
\begin{tabular}{|r|r|c|c|c|c|}
\hline
$\alpha$ & $R$ & $\bm$ & $\bs$ & ergodic & time series \\
\hline
\hline
2.5 & 1 & $1.95$ & no scaling & yes & random \\
\hline
2.5 & 500 & $1.96$ & $1.95$ & no & quasi-periodic \\
\hline
\hline
0.5 & 1 & no scaling & no scaling & yes & random \\
\hline
0.5 &100 & $1.60$ & $1.57$ & yes & random \\
\hline
\hline
0 & 1000 & $1.52$ & $1.51$ & yes & random \\
\hline
\hline
10 & 1 & $1.96$ & no scaling & yes & random \\
\hline
1.5 & 1 & $1.72$ & no scaling & yes & random \\
\hline
1 & 1 & no scaling & no scaling & yes & random\\
\hline
\hline
10 & 500 & $2.02$ & $2.01$ & no & quasi-periodic\\
\hline
2 & 500 & $1.84$ & $1.82$ & no & quasi-periodic\\
\hline
1 & 500 & $1.63$ & $1.62$ & no & quasi-periodic \\
\hline
0.9 & 500 & $1.62$ & $1.60$ & no & mode-switching \\
\hline
\end{tabular}
\end{center}
\vspace{-1pc}
\caption{\label{tab:BValues} Summary of the behavior of the \BK\ model for several values of $R$ and $\alpha$.
For $\alpha \gtrsim 1$, $P(M)$ exhibits power law behavior for all $R$
studied with the exponent $\bm \approx 2$. Power law behavior of $P(s)$
is found only for $R \gtrsim 100$; the corresponding exponent is denoted as $\bs$.
For $\alpha \lesssim 1$, $P(M)$ and $P(s)$
exhibit \mf\ scaling with slope $\bb \approx 1.5$ for $R \gtrsim 100$.
Ergodicity is discussed in Sec.~\ref{sec:Metric}, and the time series of the stress is 
discussed in Sec.~\ref{subsect:TimeSeries}.}
\end{table}

To compare our results more directly with the \ca\ models, we compute $P(s)$, the distribution of the number of blocks 
in an event (including multiple failures). As shown in Fig.~\ref{fig:PS.a}, there is no power law behavior for $R = 1$ 
and all values of $\alpha$ studied in contrast to the power law behavior of $P(M)$. This result is similar to 
that observed in Ref.~\cite{CarlsonPRA91}. Our results for $P(s)$ with $R=500$ are shown in Fig.~\ref{fig:PS.b} 
for the same values of $\alpha$ as in Fig.~\ref{fig:PS.a}. We see that for $R =500$, $P(M)$ and $P(s)$ display similar 
power law behavior with the slope $\bb \approx 2$ for $\alpha \approx 2.5$ and $\bb \approx 1.5$ for $\alpha \lesssim 1$
(see Table~\ref{tab:BValues}). As expected, the power law behavior of $P(s)$ and $P(M)$ does not hold for large $s$. These large scale characteristic events are also observed in the \ca\ models and become less probable as $R$ is increased~\cite{JunThesis, santefe,anghel}.

The scaling behavior for $\alpha = 3$, 4, 5 and 10 is similar to that for $\alpha = 2.5$.
We conclude that the generalized \BK\ model exhibits Gutenberg-Richter scaling with a \mf\ exponent of
1.5 for small $\alpha$ and large $R$. Different scaling behavior is found for $\alpha \gtrsim 2$
and all values of $R$ studied.

\section{\label{subsect:MeanSlip}The Mean Slip, Stress Drop, and the Number of Failures}

\begin{figure}[t] % fig5
\begin{center}
\subfigure[\label{fig:meanSlip.a}]{
\includegraphics[scale=0.60,angle=0]{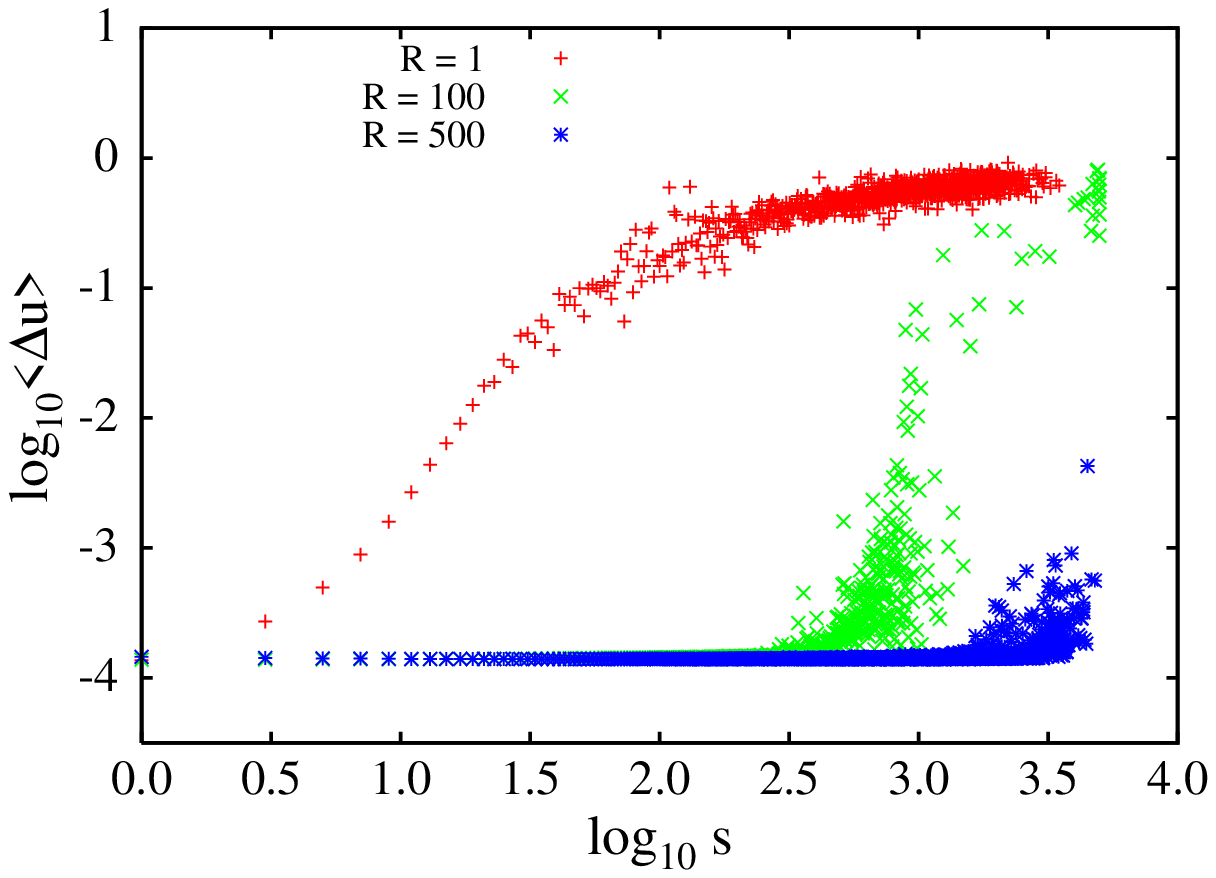}}
\subfigure[\label{fig:meanSlip.b}]{
\includegraphics[scale=0.60,angle=0]{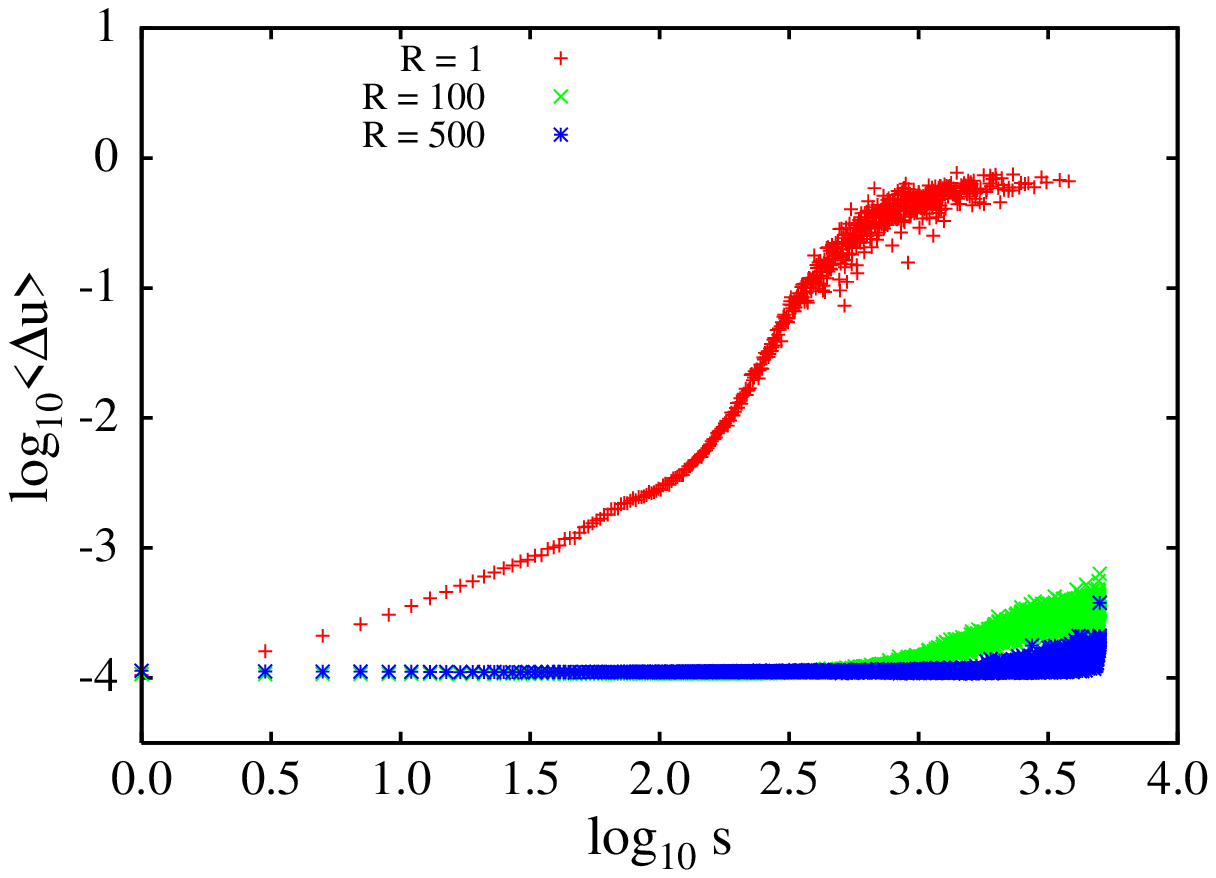}}
\end{center}
\vspace{-1pc}
\caption{\label{fig:meanSlip}(Color online) Log-log plot of the mean displacement (slip) $\du$ of a
block as a function of the number of blocks $s$ in an event for (a) $\alpha=2.5$ and (b) $\alpha=0.5$. Each block
is counted only once even if it fails multiple times. The mean slip of a block becomes independent of $s$
over a wide range of $s$ for $R \gg 1$, and the range scales with $R$. For $R \gg 1$, a characteristic 
event for $\alpha = 2.5$ has a much larger mean slip than that for $\alpha = 0.5$.}
\end{figure} 

In the CA models~\cite{RundleJSP91,OFCModelPRL92,GrassbergerPRE94,KleinPRL97,
FergusonPRE99} the stress on a block after it fails decreases to its 
residual stress, which is chosen to be either the same for all blocks, or if noise is introduced, is the same on
the average. In the \RJB\ model~\cite{RundleJSP91}, 
each failed block is displaced by an amount corresponding to the decrease of its stress. 
For large $R$ the stress on a block at failure approaches the failure threshold and therefore 
all blocks in an event are displaced the same amount. Hence, the moment
$M$ of an event and $s$, the total number of failed blocks in an event, 
are proportional for the \lr\ Rundle-Jackson-Brown model~\cite{KleinPRL97}, and the scaling behavior of $P(M)$ and
$P(s)$ becomes identical. In addition, in the \mf\ limit, a 
site fails (slips) only once during an event~\cite{KleinPRL97,FergusonPRE99}. 
These conditions are assumed by the coarse-graining theory of the long-range \RJB\ model~\cite{KleinPRL97} 
and have been verified by computer simulations~\cite{FergusonPRE99}. 
We check here if these assumptions hold for the \BK\ model for sufficiently large $R$.

Figure~\ref{fig:meanSlip} shows $\du$, the mean displacement or slip of a
block during an event, as a function of $s$. Each block is counted only once even if it slips multiple times. The behavior of $\du$ as a function of $s$ is similar for $\alpha=2.5$ and $\alpha=0.5$, and the range of 
$s$ over which $\du$ is independent of $s$ increases with the interaction
range $R$ for all values of $\alpha$ studied. The implication of this independence is that each block slips the same amount. Note that for $R=1$ there is no range of $s$ over which $\du$ is independent of $s$, even though $P(M)$ exhibits
power law behavior for small $M$ and $\alpha=2.5$.
Also note that the mean slip of a block in a characteristic event increases with $\alpha$ for fixed $R$. 

\begin{figure}[t] %fig6
\begin{center}
\subfigure[\label{fig:n2mmt.a}]{
\includegraphics[scale=0.55,angle=0]{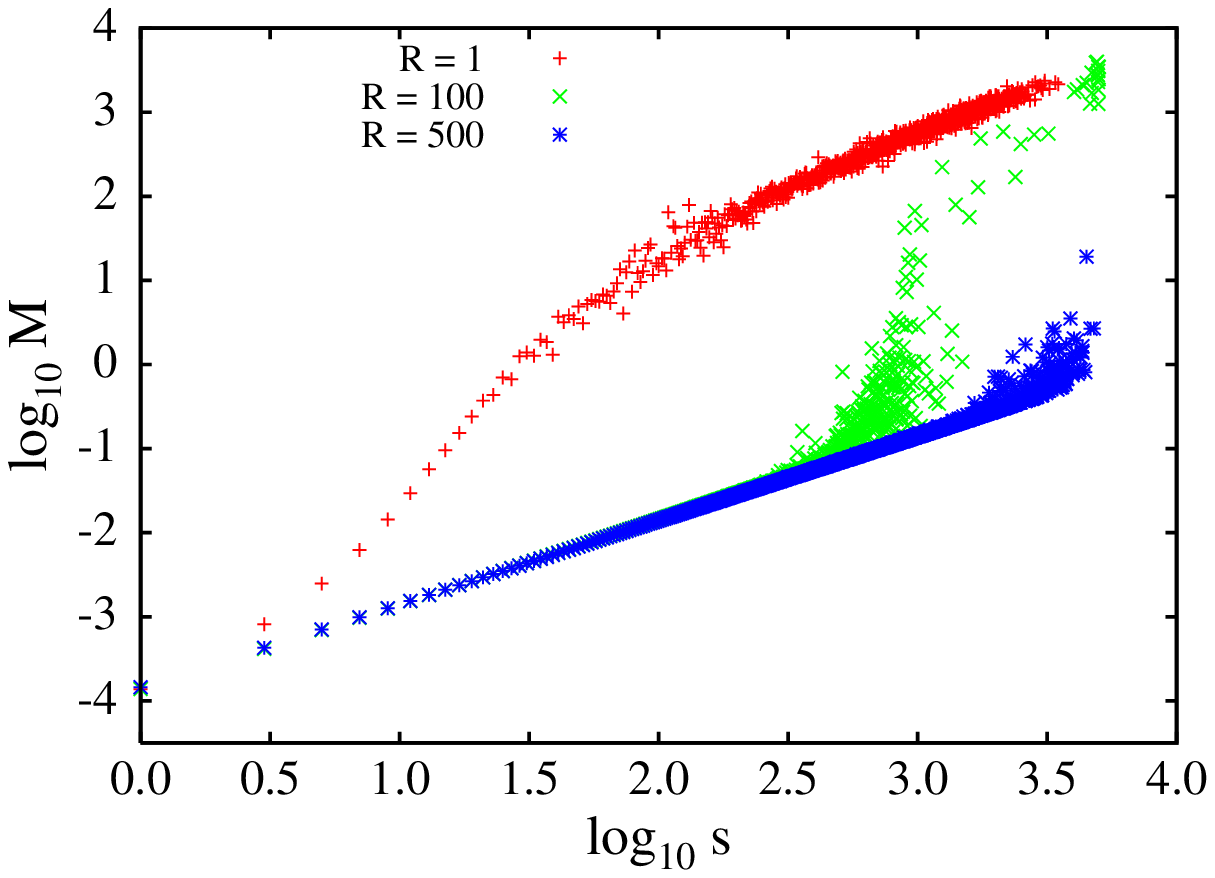}}
\subfigure[\label{fig:n2mmt.b}]{
\includegraphics[scale=0.55,angle=0]{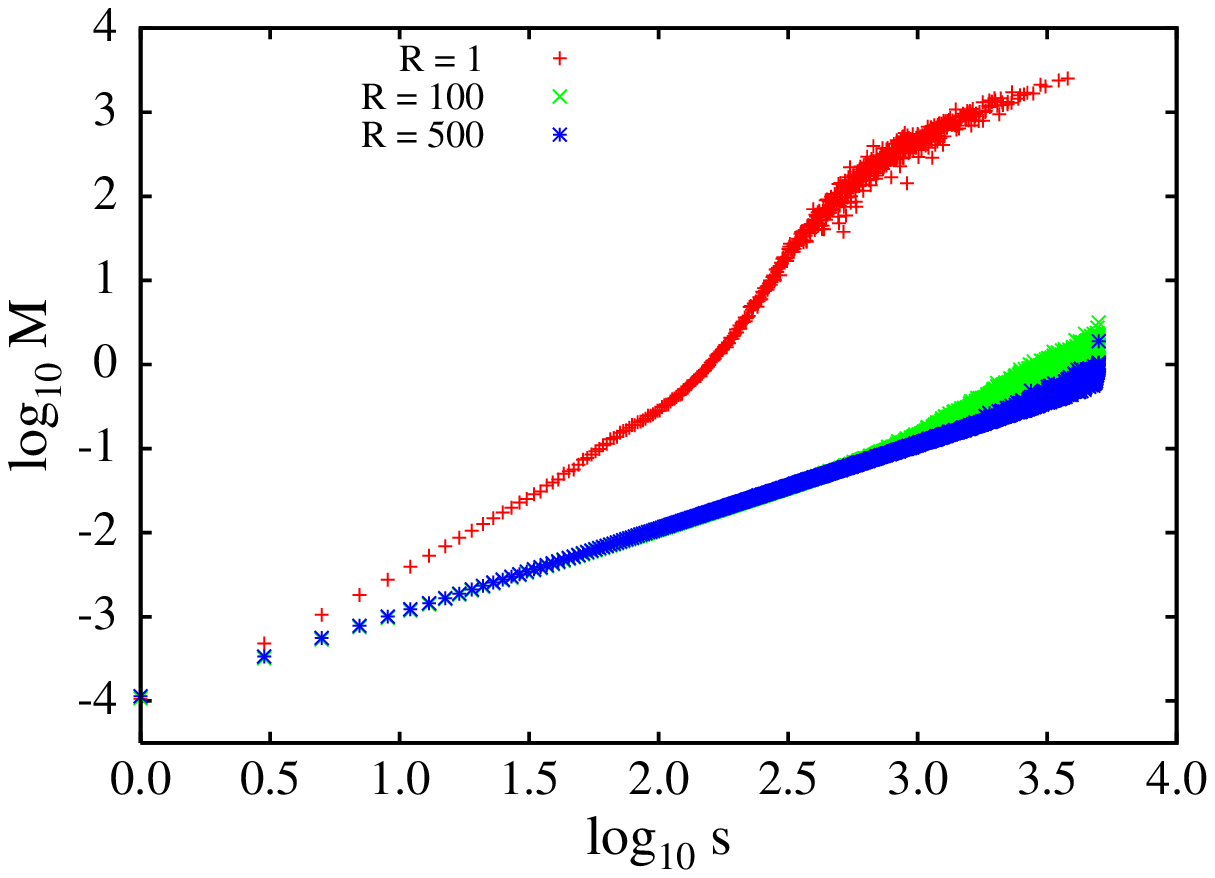}}
\end{center}
\vspace{-1.0pc}
\caption {\label{fig:n2mmt}(Color online) Log-log plot of the moment $M$ as a function of
$s$, the number of blocks in an event for (a) $\alpha=2.5$ and (b) $\alpha=0.5$. The range of $s$ for which
$M \propto s$ increases as $R$ is increased. A log-log plot is used to show 
the linear region more clearly.}
\end{figure}

Figure~\ref{fig:n2mmt} displays the moment $M$ as a function of $s$. As expected, the range of
$s$ for which $M \propto s$ increases with $R$ because the range of $s$ for which the mean displacement is constant increases with $R$. 

\begin{figure}[b] %fig7
\begin{center}
\subfigure[\label{fig:meanSdrop.a}]{
\includegraphics[scale=0.60,angle=0]{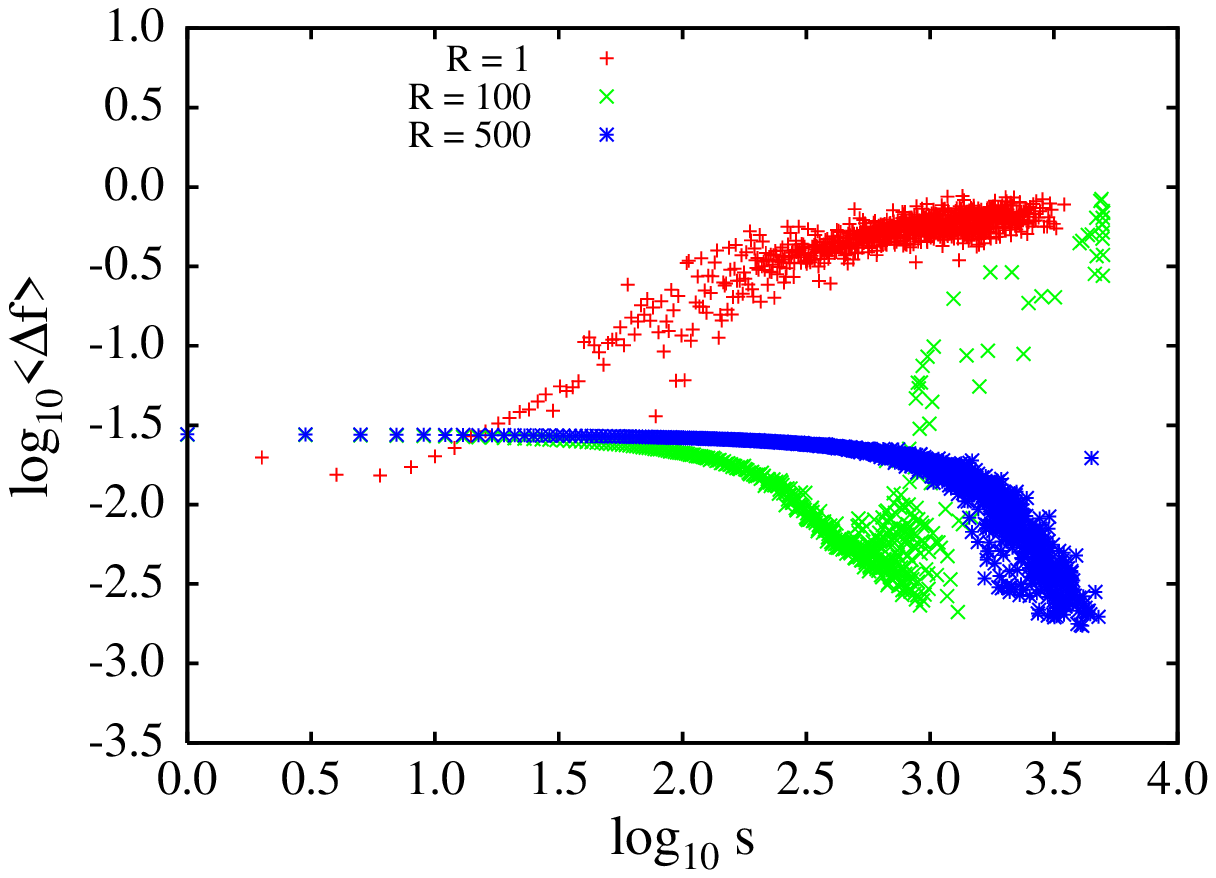}}
\subfigure[\label{fig:meanSdrop.b}]{
\includegraphics[scale=0.60,angle=0]{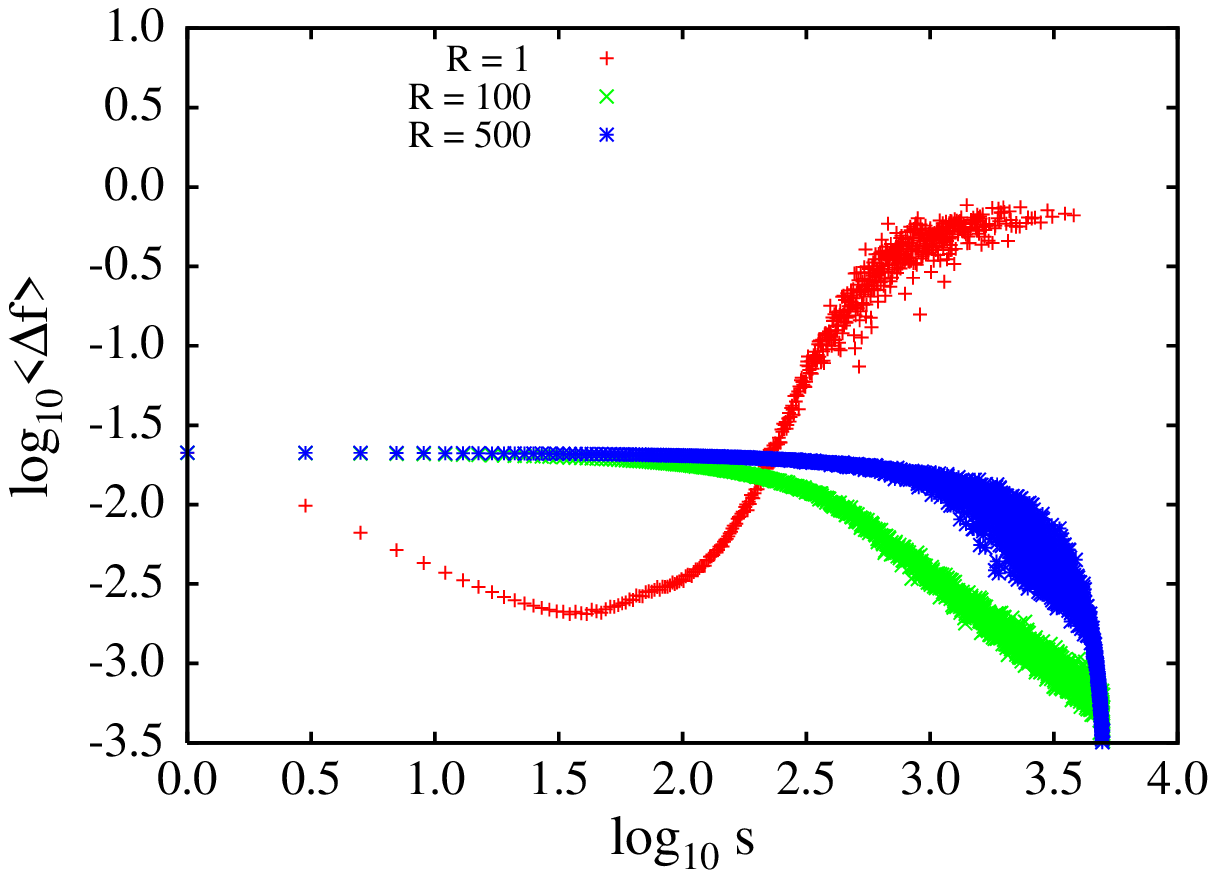}}
\end{center}
\vspace{-1pc}
\caption {\label{fig:meanSdrop}(Color online) The mean stress drop per block, $\df$, as
a function of $s$ for (a) $\alpha=2.5$ and (b) $\alpha=0.5$. 
The decrease is approximately equal to $\sigma = 0.01$ for small $s$. The range of $s$ over which the decrease is
independent of $s$ is proportional to the interaction range $R$. Note that the stress drop 
in a characteristic event for $\alpha = 2.5$ is a much larger 
than for $\alpha = 0.5$.}
\end{figure}

The mean decrease of the stress on a block after an event is given by 
\begin{equation}
\df = \frac{1}{s}\sum_i \Delta f_i, \label{eq:meanDropForce}
\end{equation}
where the sum is over all blocks in the event, and $\Delta f_i$
is the difference of the stress on the $i$th block before and after an event.
In Fig.~\ref{fig:meanSdrop} we see that the range of constant $\df$ scales with
$R$; $\df\approx \sigma = 0.01$ for power law (small $s$) events, independent of the 
value of $\alpha$.

The independence of the mean displacement and mean stress drop on the size
of an earthquake has been observed in real earthquakes and has been interpreted as evidence
for their self-similarity~\cite{ScholzBook02}. This independence
on the size of an earthquake is one of the
assumptions of the static crack model of earthquakes~\cite{ScholzBook02} and suggests that
the \BK\ model with \lr\ \st\ can 
capture some of the aspects of real earthquakes.

\begin{figure}[t] %fig8
\begin{center}
\subfigure[\label{fig:meanFtimes.a}]{
\includegraphics[scale=0.60,angle=0]{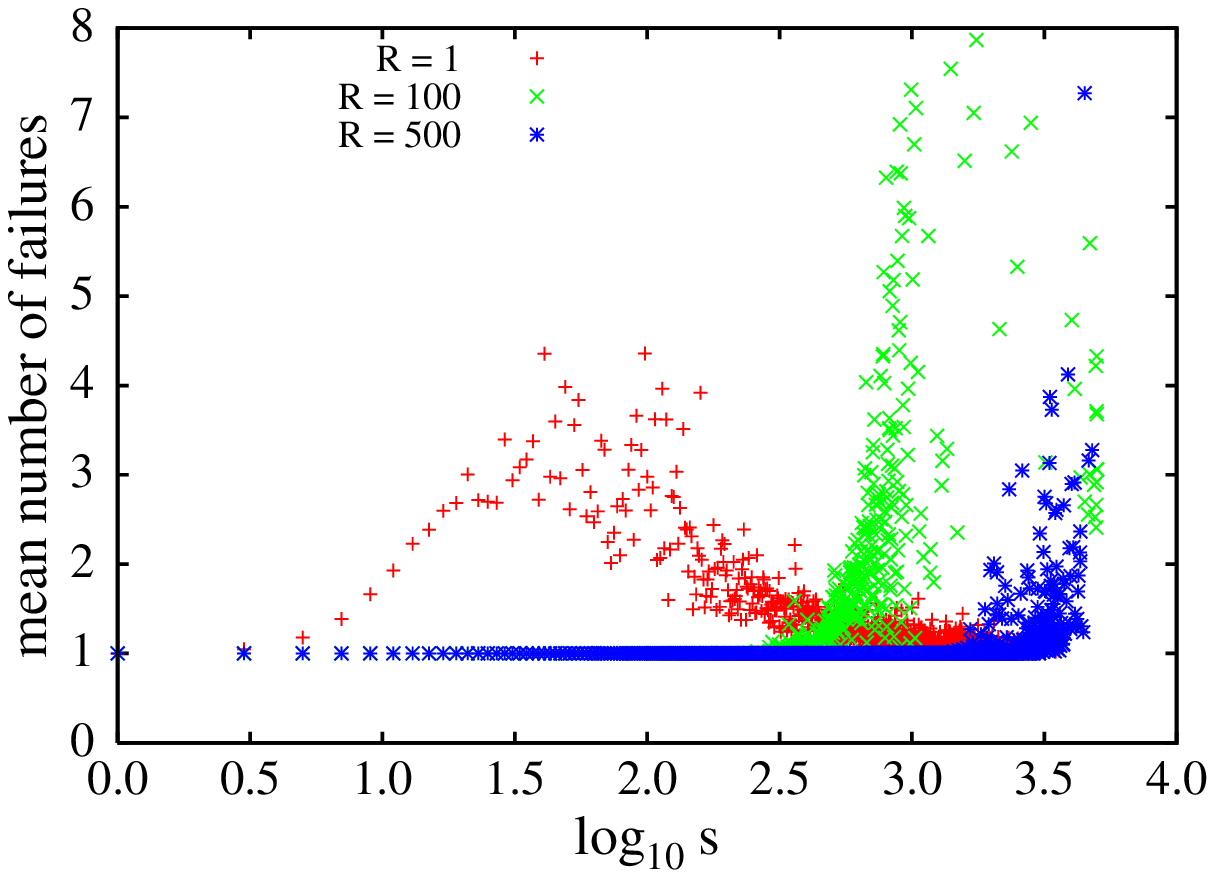}}
\subfigure[\label{fig:meanFtimes.b}]{
\includegraphics[scale=0.60,angle=0]{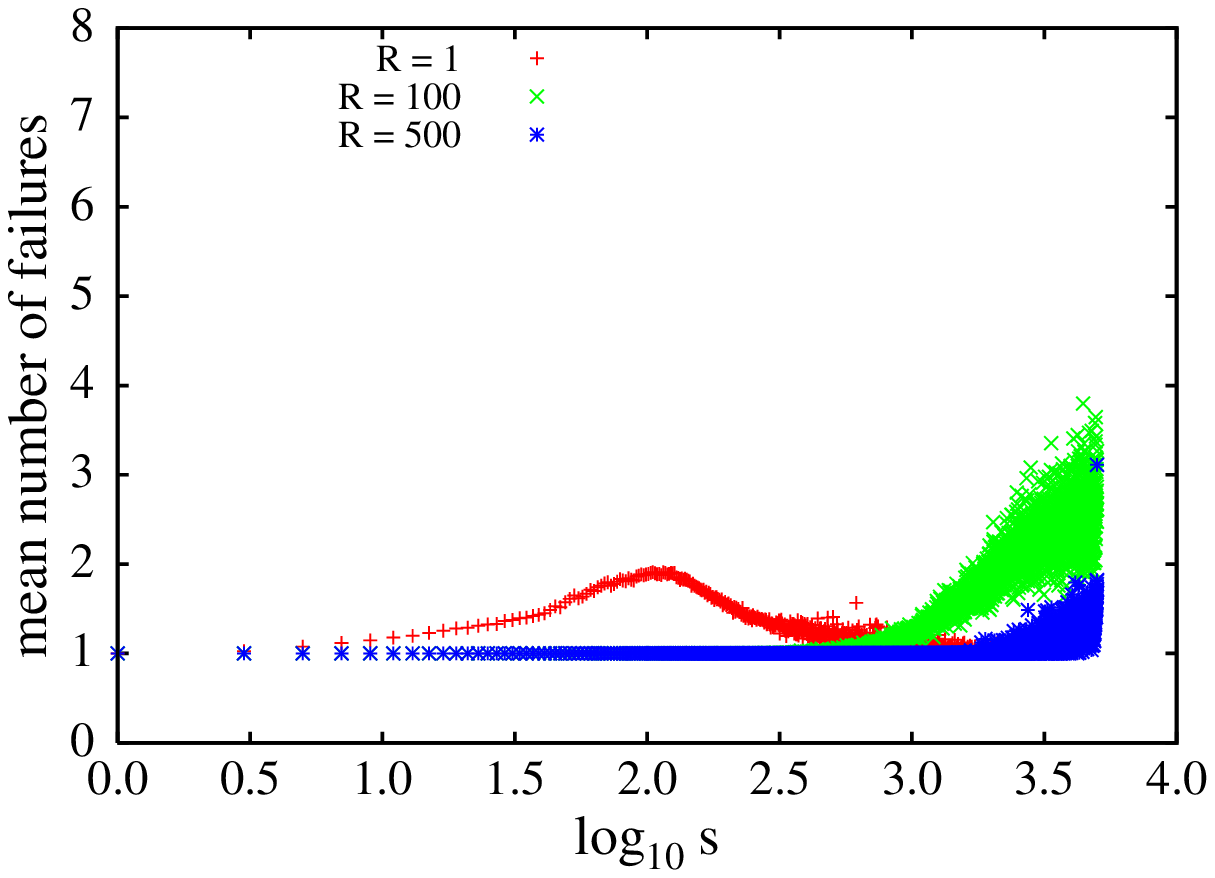}}
\end{center}
\vspace{-1.2pc}
\caption{\label{fig:meanFtimes}(Color online) The mean number of times a block fails
during an event as a function of $s$, the size of an event, for (a) $\alpha=2.5$ and (b) $\alpha=0.5$. We see that the range of $s$ over which blocks
fail only once is proportional to the interaction range $R$. 
Note that the blocks in a characteristic event fail many more times for $\alpha=2.5$ than for $\alpha=0.5$.
The curves extrapolate to 1 as $\log s \to 0$.}
\end{figure} 

Figure~\ref{fig:meanFtimes} shows the mean number of times a block fails
during an event as a function of $s$. We see
that the range of $s$ over which a block fails only once scales
with $R$; that is, multiple failures occur only for nonscaling events that are
larger than $2R$, the total number of neighbors of a block. This behavior is independent
of $\alpha$. We conclude that in the limit $R \to
\infty$, there are no blocks with multiple failures, consistent with the assumption
made in the coarse-graining description of the 
\RJB\ model in the \mf\ limit~\cite{KleinPRL97,FergusonPRE99}.

\section{\label{sec:Metric}Ergodicity and the Stress Metric}

\begin{figure}[b] %fig9
\begin{center}
\includegraphics[scale=0.71]{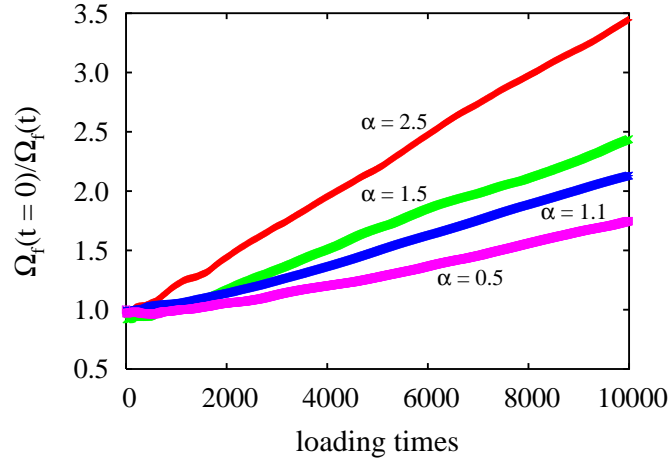}
\vspace{-1pc}
\caption{\label{fig:SMetric-alR1}(Color online) The inverse stress metric versus the
number of substrate updates (loading times) for $R=1$.
%From top to bottom the values of $\alpha$ are 2.5, 1.5, 1.1, and 0.5.
The system is ergodic for all the values of $\alpha$ studied, and the slope 
is an increasing function of $\alpha$ (see Table~\ref{tab:BValues} for more values of $\alpha$).}
\end{center}
%\vspace{-2.0pc}
\end{figure}

\begin{figure}[t] %fig10
\begin{center}
\subfigure[\label{fig:StrMetric.a}]{
\includegraphics[scale=0.60]{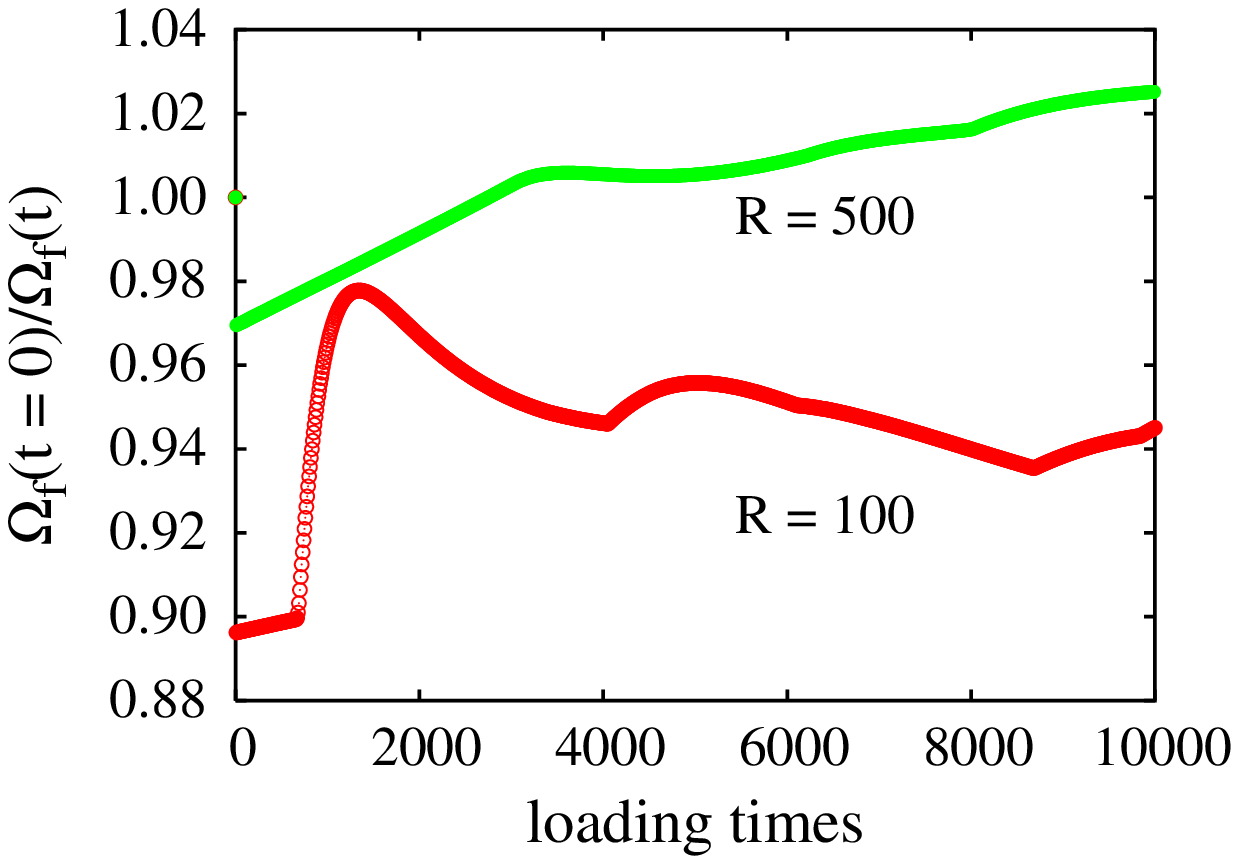}}
\subfigure[\label{fig:StrMetric.b}]{
\includegraphics[scale=0.60]{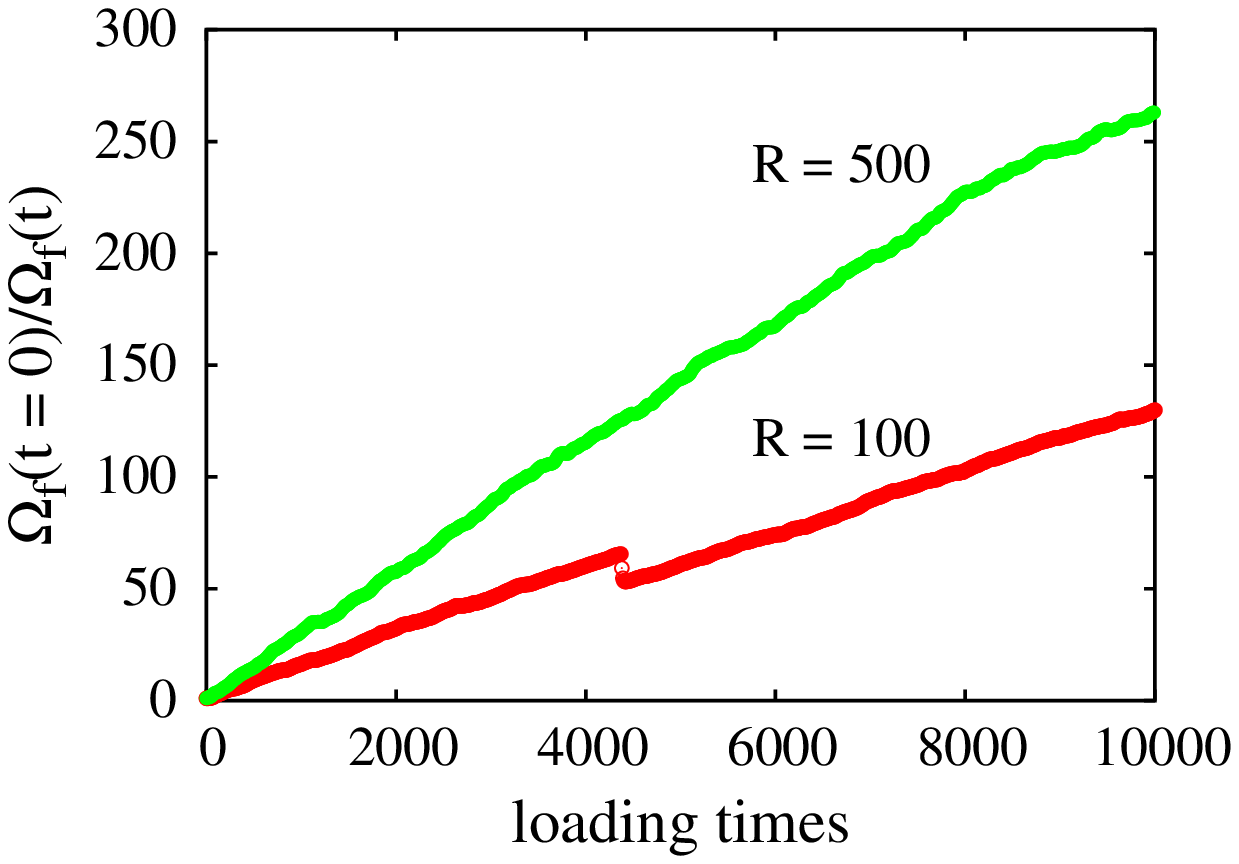}}
\subfigure[\label{fig:StrMetric.c}]{
\includegraphics[scale=0.60]{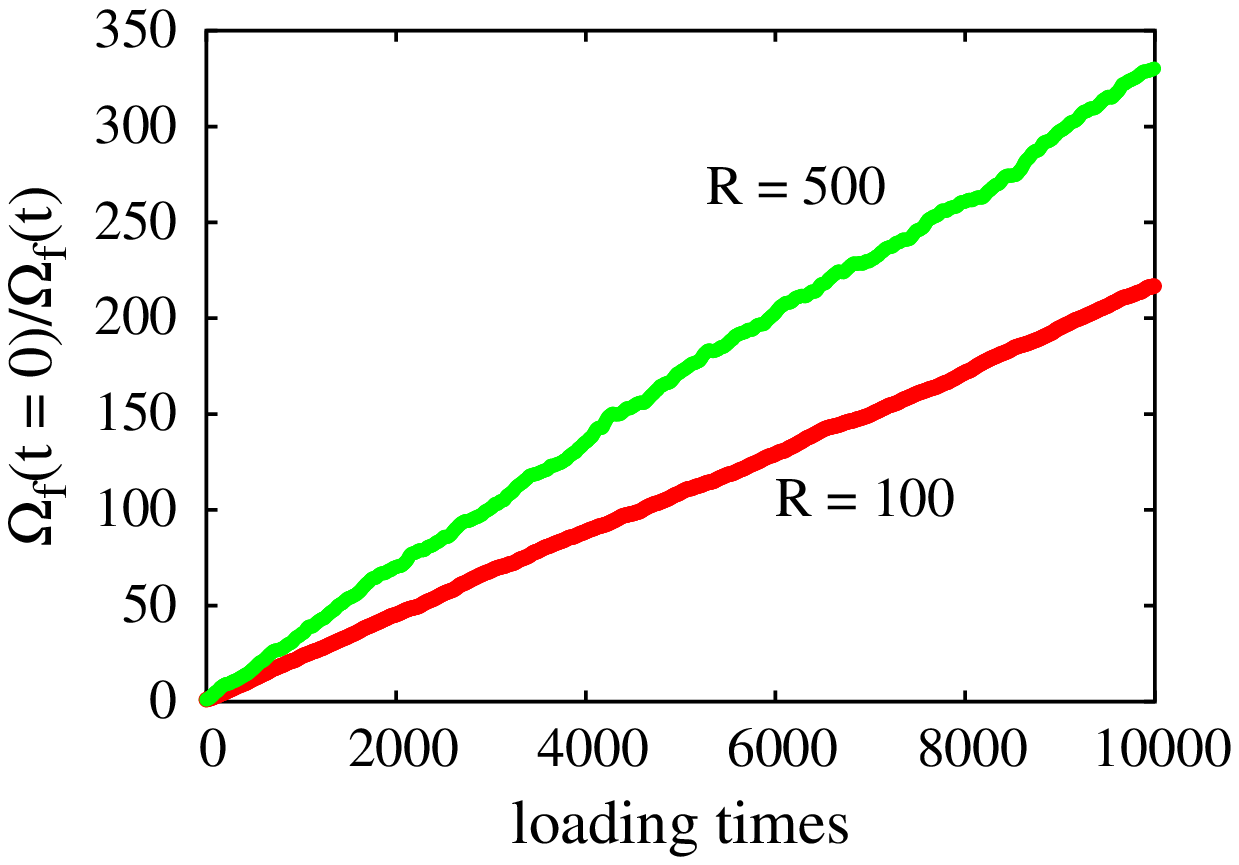}}
\vspace{-.1pc}
\caption{\label{fig:StrMetric}(Color online) The normalized inverse stress metric versus the number of substrate updates (loading times) for $R=100$ and 500 and (a) $\alpha=2.5$, (b) $\alpha=0.5$, and (c) $\alpha=0$. 
Note the different vertical scales. The system is nonergodic for
larger values of $R$ for $\alpha=2.5$. For
$\alpha=0.5$ and $\alpha=0$, the slope of $\Omega(0)/\Omega(t)$ becomes larger as
$R$ is increased. The system exhibits punctuated ergodicity for
$\alpha = 0.5$ and $R=100$, similar to the behavior
of the \lr\ CA models~\cite{FergusonPRE99}. See Ref.~\cite{JunThesis} 
for more values of $\alpha$ and $R$.}
\end{center}
%\vspace{-2.0pc}
\end{figure}

We characterize the nature of ergodicity in the \BK\ model by determining the
metric $\Omega_f(t)$ of the stress~\cite{TMErgodicity}. We take $f_i(t)$ to be a
quantity associated with block
$i$ and define
\begin{align}
\label{tm1}
{\overline f}_i(t) &= {1\over
t}\!\int_{0}^{t}\! f_i(t^{\prime})\,dt^{\prime} \\
\label{tm2}
\lb f(t) \rb &= {1\over N} \sum_{i = 1}^{N}\overline{f}_i(t),
\end{align}
and the metric
\begin{equation}
\Omega_{f}(t)= {1\over N}\sum_{i = 1}^{N}\big\lbrack {\overline f}_i(t) - \lb f(t) \rb\big\rbrack^{2}. \label{tm3}
\end{equation}
If the system is ergodic, 
$\Omega_{f}(t)\propto 1/t$~\cite{TMErgodicity}. Because the metric studied
in the CA models is the stress metric~\cite{FergusonPRE99}, we choose $f_j$ to be the stress on block $j$ just after an
event.

In Fig.~\ref{fig:SMetric-alR1} we show the normalized inverse stress metric 
$\Omega_f(0)/\Omega_f(t)$ for $R=1$ and different values of $\alpha$. We see
that
$\Omega_f(0)/\Omega_f(t)$ increases linearly with $t$. The mixing time $\tau$ can be defined by the relation $\Omega_f(0)/\Omega_f(t)=t/\tau$. We see that the mixing time $\tau$ decreases with increasing $\alpha$~\cite{TMErgodicity}.
We conclude that the nearest-neighbor ($R=1$) \BK\
model is ergodic for all values of $\alpha$ studied.

The system exhibits qualitatively different behavior for larger
values of $R$. For $\alpha=2.5$ the system is nonergodic for $R \gtrsim 100$ during our observation time. In contrast, for $\alpha \lesssim 1$ 
the system remains ergodic as $R$ is increased and the mixing time $\tau$ decreases with increasing $\alpha$ as for $R=1$ (see Figs.~\ref{fig:StrMetric.b} and \ref{fig:StrMetric.c}). 
Note that for $\alpha = 0.5$ and $R = 100$, the system displays punctuated ergodicity during 
our observation time, similar to the behavior of the \lr\ CA models~\cite{FergusonPRE99}. 

Punctuated ergodicity has been observed in the Southern California 
fault system for some coarse graining conditions~\cite{TiampoPRL03}. 
Ergodicity can also be recovered for a simple, 
far-from-equilibrium system with underlying chaotic dynamics serving as a temperature
bath~\cite{EgolfErgodicity}. Our results suggest the possibility of
both ergodic faults and nonergodic faults.

\section{The Time Series of the Stress}
\label{subsect:TimeSeries}

\begin{figure}[h] %fig11
\begin{center}
\subfigure[]{
\includegraphics[scale=0.55]{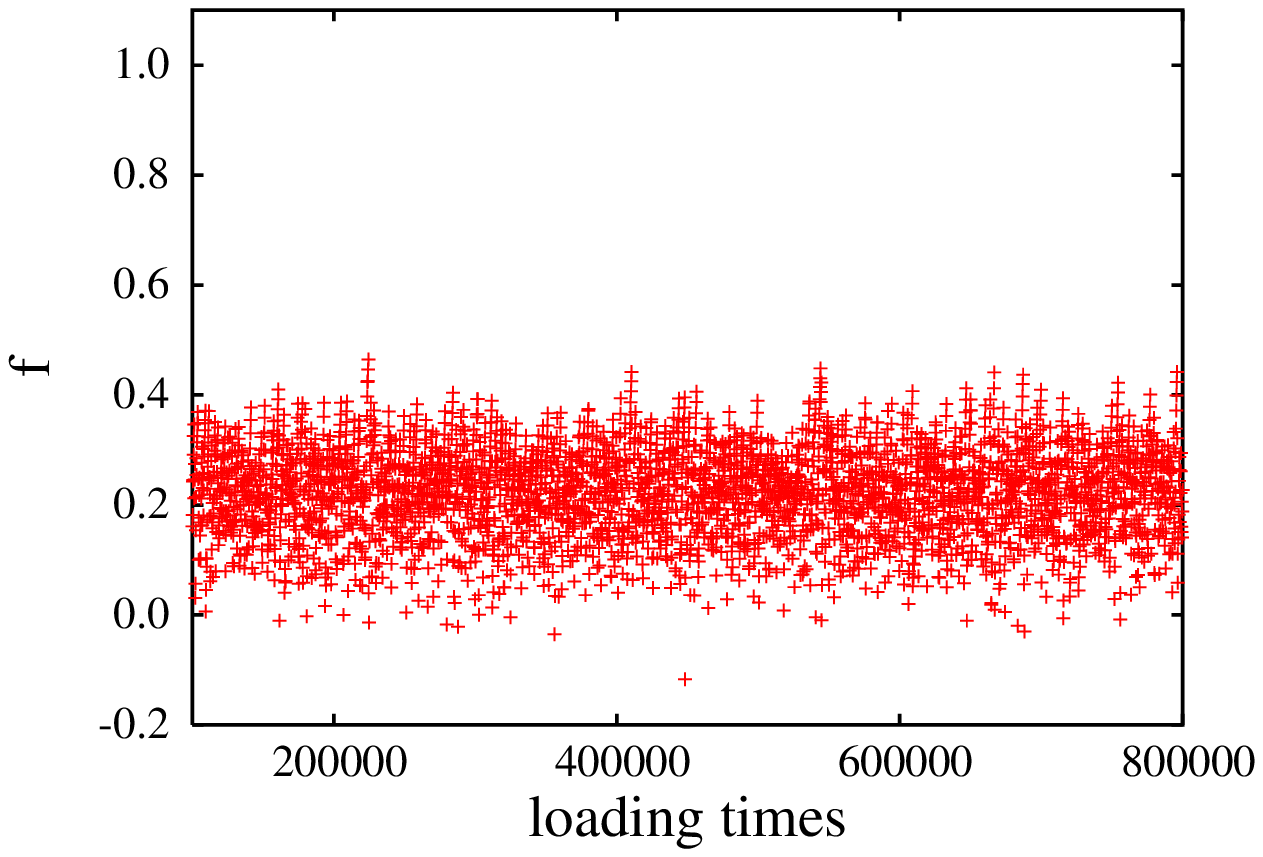}}
\subfigure[\label{fig:11b}]{
\includegraphics[scale=0.55]{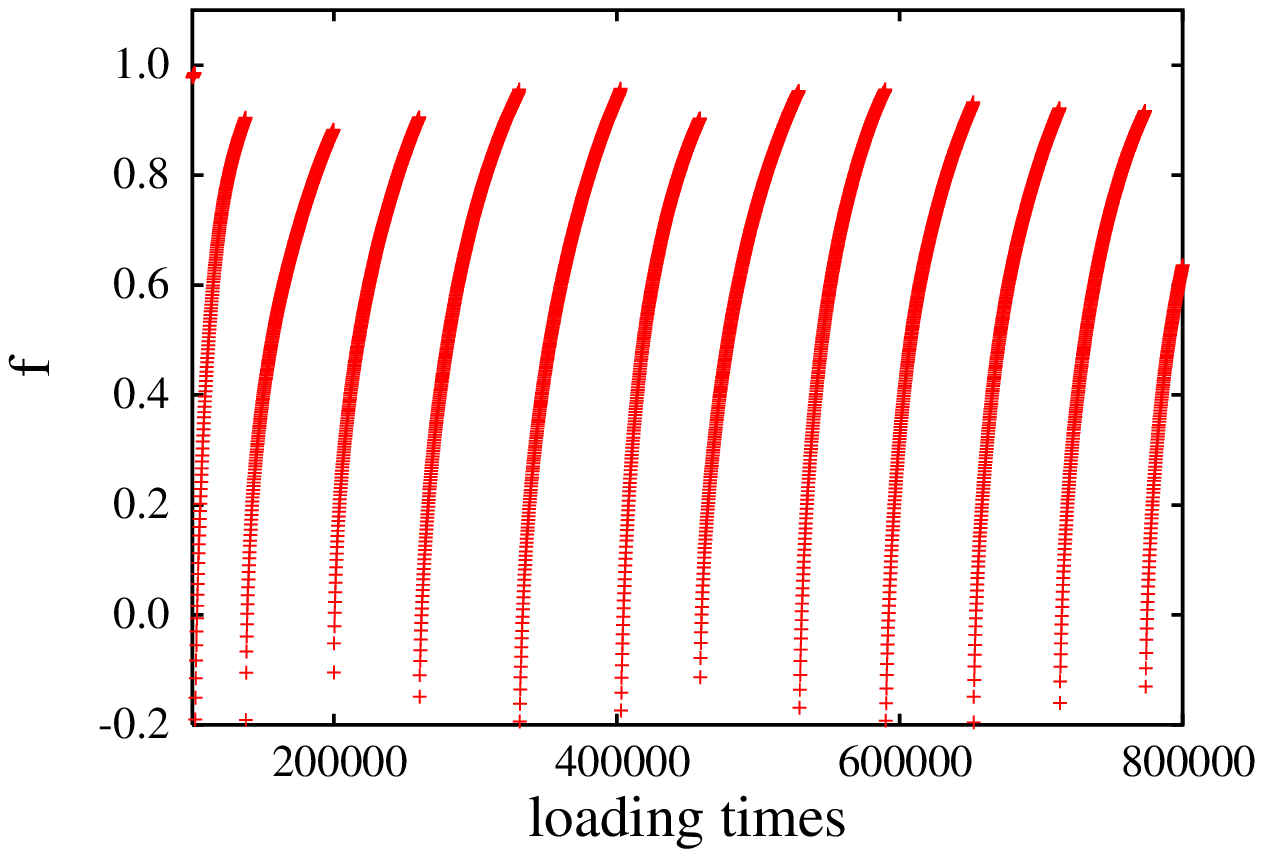}}
\subfigure[\label{fig:11c}]{
\includegraphics[scale=0.55]{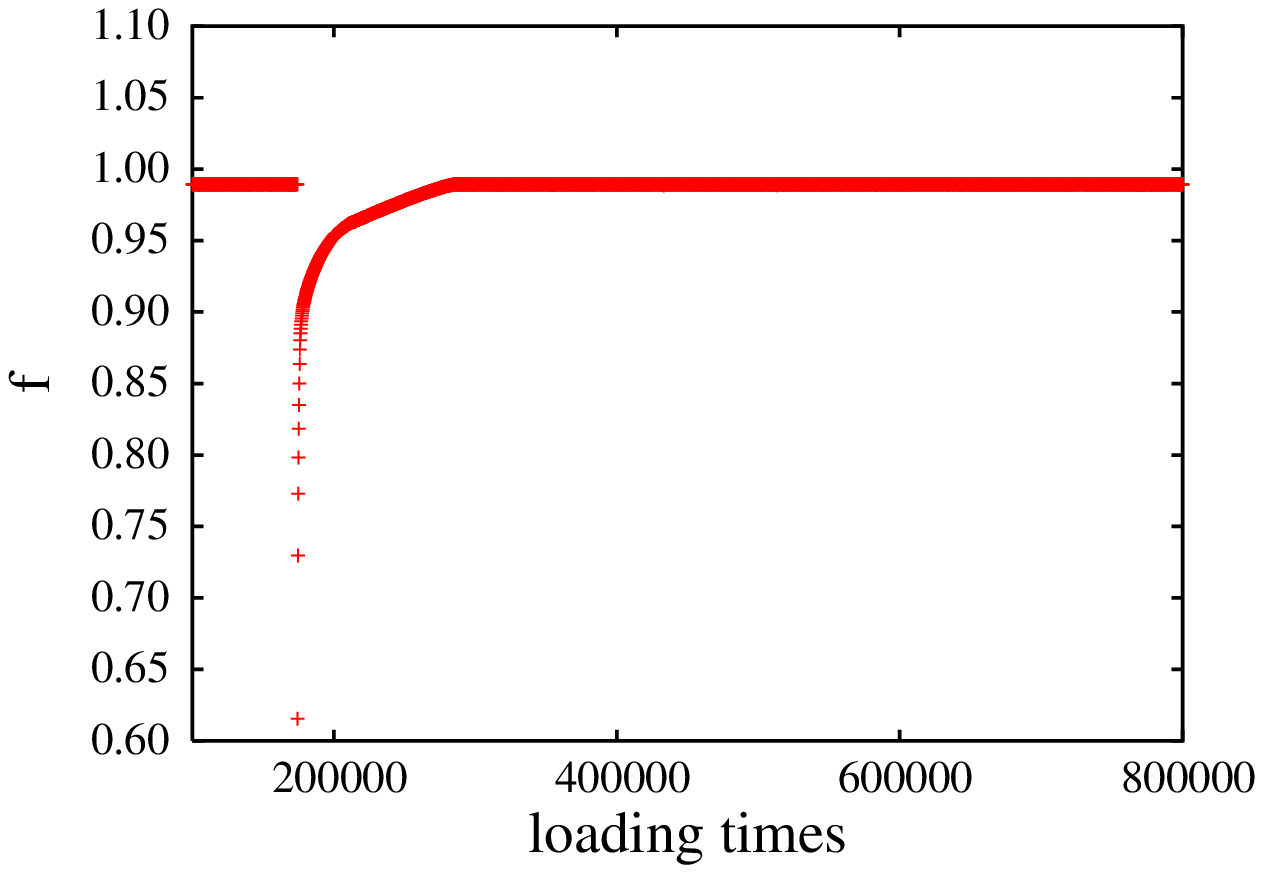}}
\subfigure[\label{fig:11d}]{
\includegraphics[scale=0.55]{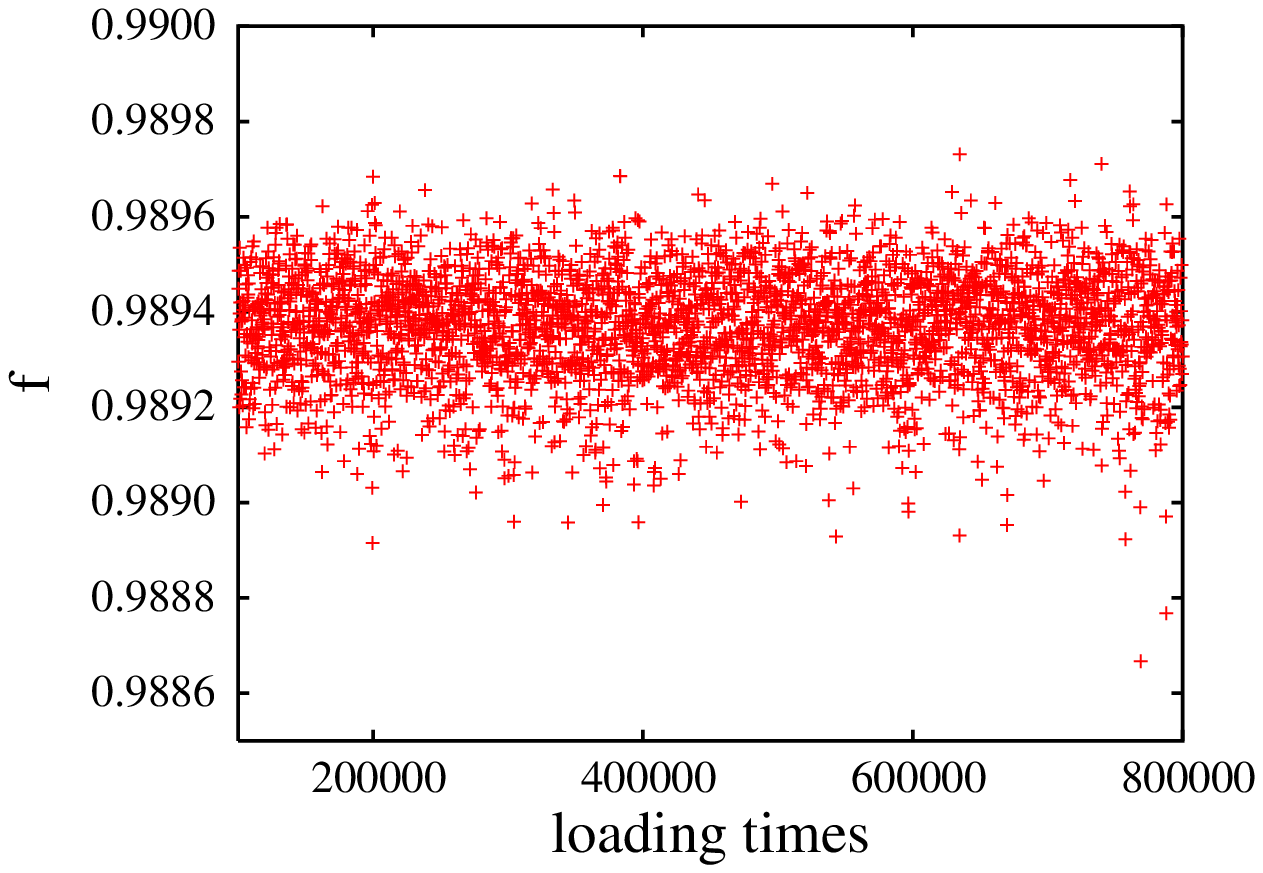}}
\caption{\label{fig:StrTal25+0} Time series of the stress $f$ per block as a
function of substrate plate updates (loading times) for (a) $R=1$, $\alpha = 2.5$, (b) $R=500$, $\alpha = 2.5$, (c) $R=100$, $\alpha = 0.5$, and (d) $R=500$, $\alpha = 0.5$. For $R = 1$
the stress fluctuates between 0 and 0.4. As $R$ is increased, the time
series becomes quasiperiodic and the system is no longer ergodic. The sharp decrease in $f$ for $R = 100$ and $\alpha=0.5$ corresponds to the punctuated ergodicity seen in Fig.~\ref{fig:StrMetric.b}. As $R$ is increased, no quasiperiodic behavior is observed, and $f$ fluctuates about a high stress state close to $1-\sigma$.}
\end{center}
%\vspace{-1pc}
\end{figure}

To understand the behavior of the stress metric, we plot the times series of
the stress
$f$ per block just after an event (see 
Fig.~\ref{fig:StrTal25+0}). For
$\alpha=2.5$ and
$R=1$, the times series fluctuates between 0 to 0.4. This behavior is
consistent with the observed ergodicity of the system (see
Fig.~\ref{fig:SMetric-alR1}). As
$R$ is increased (Fig~\ref{fig:11b}), the time series becomes quasiperiodic and the period
becomes longer for larger $\alpha$~\cite{JunThesis}. For
$R = 500$, $f(t)$ is quasiperiodic with a mean of $\approx 0.4$ and
a range from $\approx -0.2$ to almost 1. This quasiperiodic time dependence from a
lower stress state to a high stress state is the origin of the nonergodicity
for
$\alpha
\gtrsim 1$ and $R \gg 1$. That is, small earthquakes accumulate stress
locally and characteristic earthquakes release the stress globally and
quasiperiodically. The periodicity of
large characteristic earthquakes for
$R=1$ and $\alpha=2.5$ was also observed in Refs.~\cite{MoriBK, MoriBK2}.
The quasiperiodic behavior in Fig.~\ref{fig:StrTal25+0} is reminiscent of the
stress versus time curves observed in laboratory experiments with
rocks~\cite{km}. 

For
$R = 1$ and $\alpha=0.5$~\cite{JunThesis}, there are intervals where the stress increases after small
events. However, many random decreases occur and the time series fluctuates 
between 0.3 and 0.8 similar to the behavior for $\alpha = 2.5$
and $R=1$. These random fluctuations are consistent with the system being
ergodic (see Fig.~\ref{fig:SMetric-alR1}). In contrast with the behavior of
the time series for
$\alpha=2.5$, no quasiperiodic behavior is observed as
$R$ is increased (see Figs.~\ref{fig:11c} and \ref{fig:11d}). 
Instead,
$f(t)$ remains in a high stress state, $\lb f \rb \approx 0.99$,
and the system remains ergodic. This behavior is also observed in the \lr\ CA
models~\cite{FergusonPRE99}. Note that for $R = 100$, the stress returns
to a relatively small value only once during the observation time, which
makes the system exhibit punctuated ergodicity as shown in
Fig.~\ref{fig:StrMetric.b}.

Increasing $\alpha$ for $R = 500$ makes the
quasiperiodic behavior of $f(t)$ better defined and increases the period~\cite{JunThesis}. For $\alpha = 0.9$ (not shown)~\cite{JunThesis} the system remains in a high stress state for some time and
exhibits mode-switching behavior similar to the long-range \OFC\ model with
a long healing time~\cite{WeatherleyPAG02}, and a cellular automaton model
of a vertical fault with dynamic weakening of cell
strengths~\cite{BenzionJGR95}. Ben-Zion et al.~\cite{BenzionEPSL99} have noted the importance 
of mode switching in understanding earthquake fault systems.

\section{Waiting-Time Distribution}
\label{subsect:WaitTDistr}

\begin{figure}[t] %fig12
\begin{center}
\subfigure[\label{fig:WTDFN5+3R1+0al25+0}]{
\includegraphics[scale=0.62]{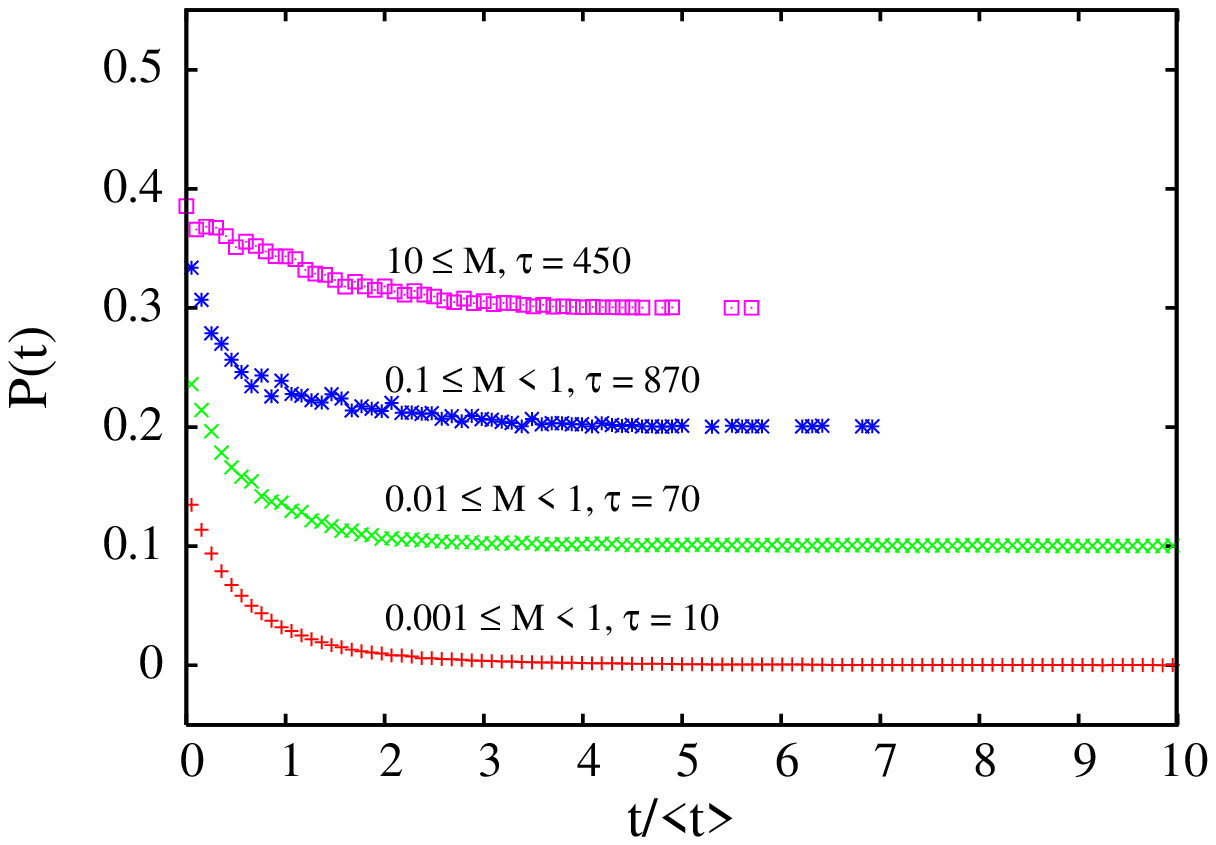}}
\subfigure[\label{fig:WTDFN5+3R5+2al25+0}]{
\includegraphics[scale=0.62]{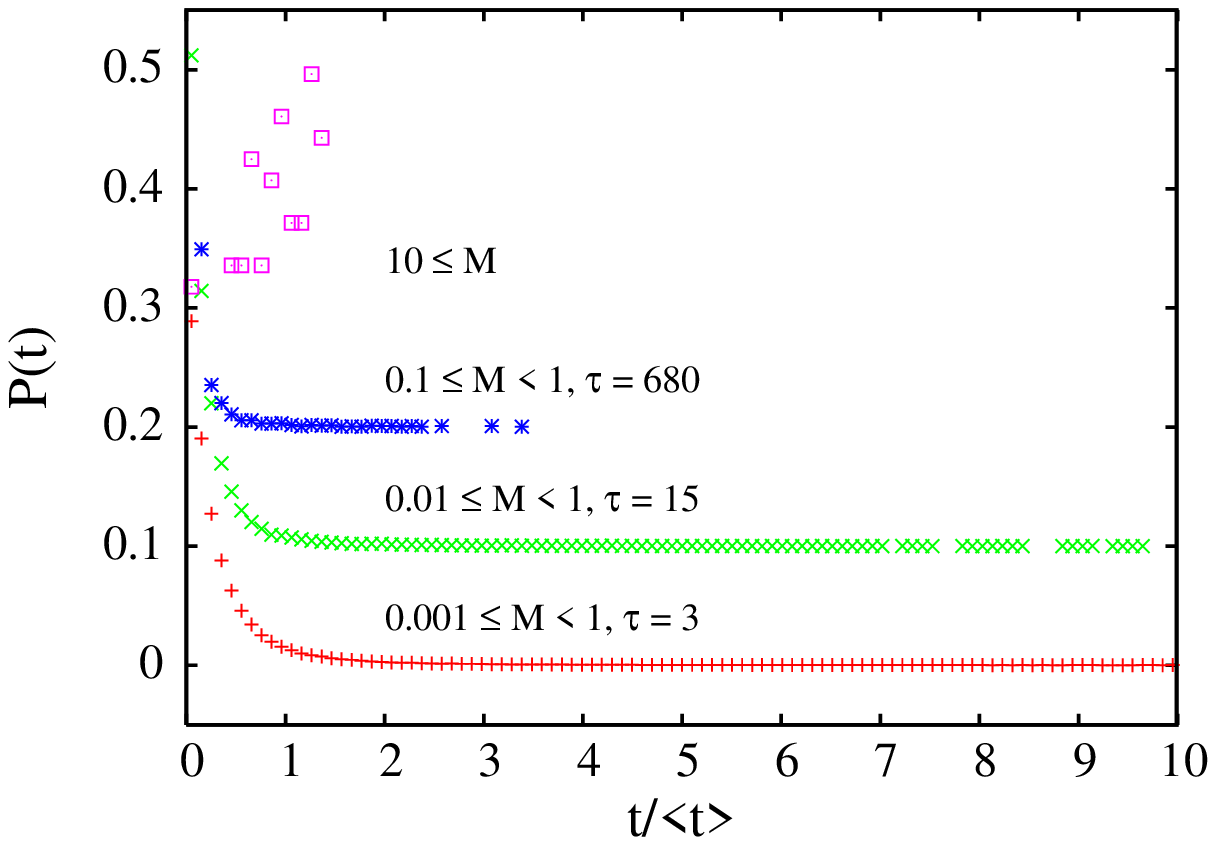}}
\subfigure[\label{fig:WTDFN5+3R1+0al50-1}]{
\includegraphics[scale=0.62]{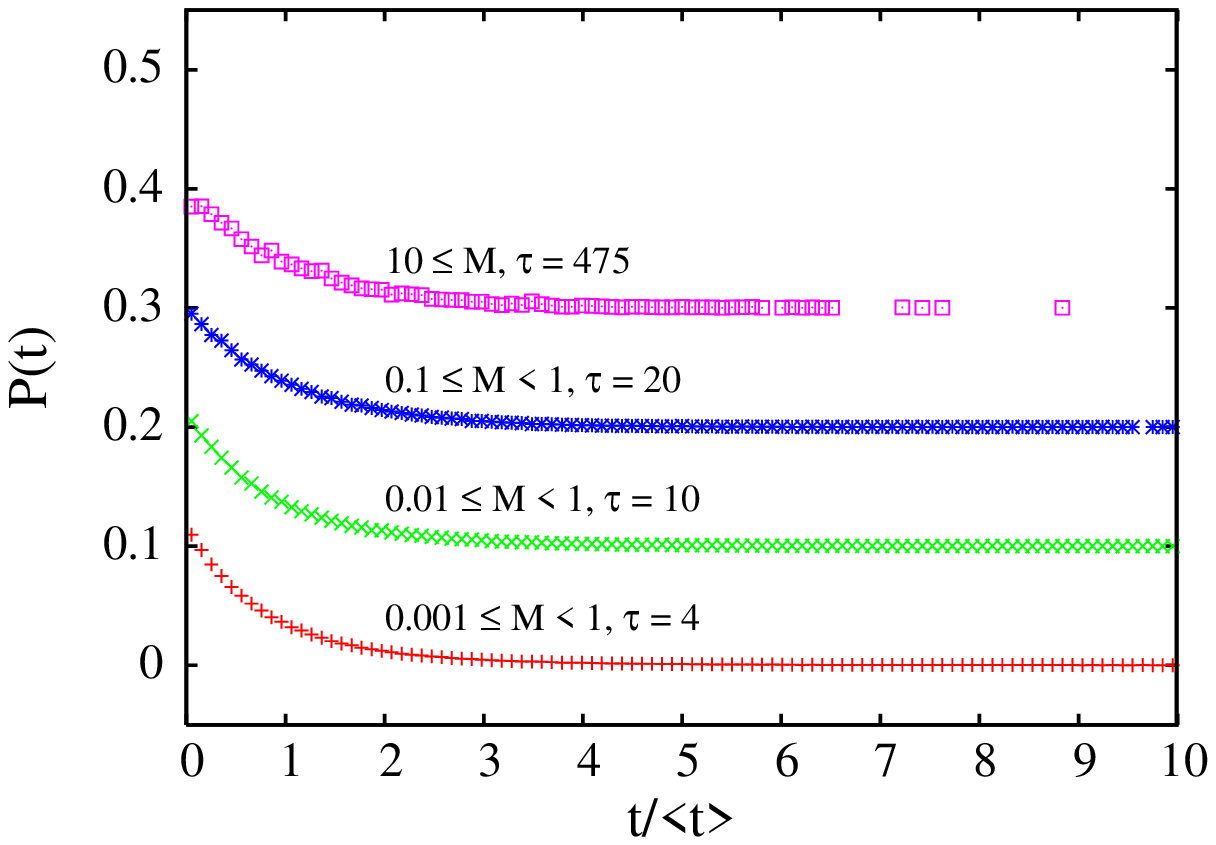}}
\subfigure[\label{fig:WTDFN5+3R5+2al50-1}]{
\includegraphics[scale=0.62]{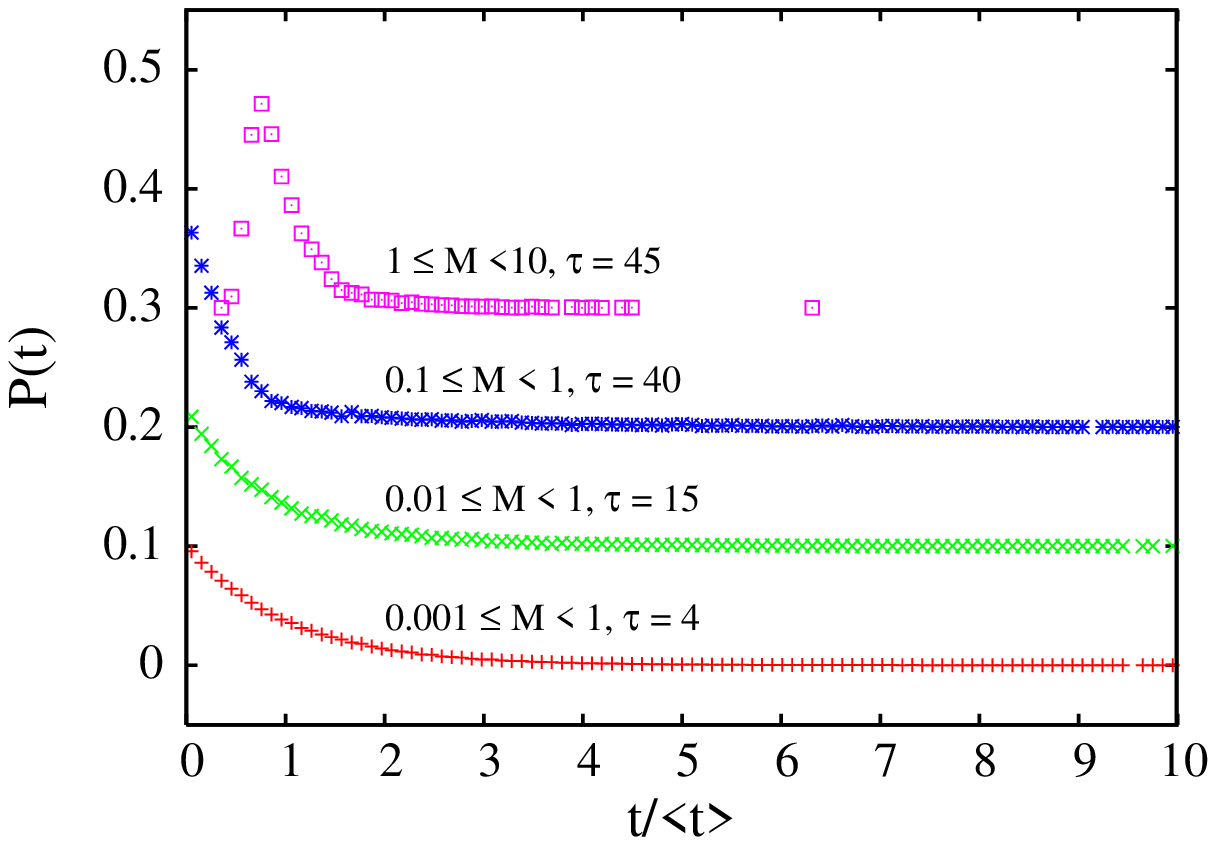}}
\end{center}
\vspace{-0.5cm}
\caption{(Color online) The waiting-time distribution $P(t)$ for (a) $R=1$, $\alpha=2.5$, (b) $R=500$, $\alpha=2.5$, (c) $R=1$, $\alpha=0.5$, and (d) $R=500$, $\alpha=0.5$. The time is scaled by the mean waiting time $\lb t \rb$. The waiting-time distributions are fitted
to the exponential form $f(t) \propto e^{-t/\tau}$ if possible. The characteristic time $\tau$ of the exponential decay increases with $M$ for the 
power law events. The waiting-time distribution
$P(t)$ is close to an exponential 
for both the scaling events and the characteristic events for $R = 1$.
For $R = 500$ the waiting-time distribution for the scaling events is close to an exponential, but
$P(t)$ for events with $M \geq 1$ is not.
For clarity, each distribution is shifted vertically by 0.1.}
\label{fig:WTDF-RN5+3al25+0}
\end{figure} 

The nature of the waiting or recurrence-time distribution $P(t)$ for events of a given range of sizes
is of current interest~\cite{YangInterEvents,CorralInterEvents,LindmanInterEvents}. Statistical data from the Southern California Fault Network
show there exists correlations, at least between large 
events~\cite{YangInterEvents,CorralInterEvents,LindmanInterEvents}. These correlations imply that 
the waiting-time distribution is not exponential. 

We assume that the substrate moves 
with a constant velocity and hence the time $t$ between two events is proportional to the number of
substrate updates. Figure~\ref{fig:WTDF-RN5+3al25+0} shows our results for $P(t)$ for $\alpha = 0.5$ and 2.5 and $R=1$ and 500.
For all combinations of $R = 1$ and $\alpha$ the waiting-time distribution for both the scaling events and 
characteristic events is close to an exponential distribution, which implies that there is little correlation between the events.
The exponential decay of $P(t)$ for $R = 1$ and $\alpha = 2.5$ has been reported in 
Refs.~\cite{MoriBK, MoriBK2}. For $R = 500$ and $\alpha=2.5$ (see Fig.~\ref{fig:WTDFN5+3R5+2al25+0}) the waiting time distribution for the characteristic events is nonexponential because the characteristic events are quasiperiodic for large $R$. Because there were no characteristic events for $R=500$ and $\alpha=0.5$ during our observation time, we computed $P(t)$ for events in the range $1 \leq M < 10$ and found that there is a maximum at $t \approx 100$ and a long exponential tail. Hence, we conclude that the large events in long-range models for $\alpha=2.5$ and $\alpha=0.5$ are correlated.

\section{summary}
\label{sec:Summary}

The \BK\ model with \lr\ \st\ is more realistic than the usual \bk\ model
with nearest neighbor interactions and is more realistic than the
\lr\ CA models because of the presence of inertia and a dynamic friction
force. Our results show that the generalized \BK\ model exhibits much
richer scaling behavior than the \ca\ models and its behavior depends on 
the nature of the velocity-weakening friction force and the range of the stress transfer~\cite{mori3}.

For nearest-neighbor interactions ($R = 1$) we verified the power law
behavior of the moment distribution $P(M)$ for $M \lesssim 1.0$ and $\alpha \gtrsim 1$ with a
scaling exponent of $\bb \approx 2$ for $\alpha \gtrsim 2$~\cite{CarlsonPRA89,CarlsonPRA91}.
Qualitatively similar results (not shown)~\cite{JunThesis} were found for $\alpha = 3$, 4, 5,
and 10 with $\ell=10$ and for $\ell=5$, 7, 10, and 20 with $\alpha=2.5$. No
scaling was found for the moment distribution for $\alpha \lesssim 1$, and
the distribution of the number of failed blocks during an event, $P(s)$,
does not scale for all values of $\alpha$ considered.

For the \lr\ model ($R \gg 1$), $P(M)$ and $P(s)$ show similar
behavior. For $\alpha \gtrsim 1$, power law behavior with an apparent
exponent of $\bb \approx 2$ as for $R =1$ was found with the additional 
presence of characteristic events. As $R$ is increased, the range of power
law behavior becomes smaller. For $\alpha \lesssim 1$, an exponent of $\bb \approx 1.5$ was obtained for $R \gg 1$. This value of $\bb$ is consistent
with that found for the \lr\ CA models and the existence of a spinodal critical point~\cite{RundlePRL95,KleinPRL97,FergusonPRE99,KleinGeoComp00}.
The probability of characteristic events for small $\alpha$ decreases rapidly with increasing $R$.

We found that the \BK\ model is ergodic for $R=1$ and all values of $
\alpha$ studied; this behavior is 
in agreement with the random fluctuations in the time series of the 
stress. 
For $R \gg 1$, the system is ergodic for $\alpha \lesssim 1$ because 
the system fluctuates about 
a high stress state. This behavior is consistent with the behavior of the \lr\ CA models~\cite
{RundlePRL95,KleinPRL97,FergusonPRE99}
and the observation of the Southern California fault system~\cite
{TiampoPRL03}. 
The system becomes nonergodic for $R \gg 1$ and $\alpha \gtrsim 1$ due to 
the quasiperiodic behavior
of the stress. For $\alpha = 0.9$ and $R = 500$, we
found mode-switching between quasiperiodic behavior and fluctuations
around a high-stress state, similar to that found in other
models~\cite{WeatherleyPAG02,BenzionJGR95,BenzionEPSL99}. The
exponential fits to the distribution of
waiting times for the scaling events for the values of $\alpha$ and $R$ studied
implies that there is no correlation between these events, which is inconsistent
with the statistical data from real fault network systems~\cite{helm}. However,
large events such as characteristic events are correlated.

Our simulation results suggest that there exists two scaling regimes with qualitatively
different behavior, one of which ($\alpha \rightarrow 0$ and $R \gg 1$) is
consistent with an equilibrium spinodal critical point and a \mf\
exponent of $\bb = 3/2$, similar to the \lr\ CA models and the \nmf\
picture of spinodal nucleation~\cite{RundlePRL95,KleinPRL97,KleinGeoComp00}.
The nature of the other scaling regime with $\bb \approx 2$ for $\alpha \gtrsim 1$ is not well understood~\cite{CarlsonPRA89,CarlsonPRA91,Nagelgroup}. The apparent dependence of
the scaling exponent $\bb$ on $\alpha$ for $1 \lesssim \alpha \lesssim 2.5$ suggests that the
interpretation of this scaling regime in terms of a dynamical critical
point must be viewed with caution and that larger system sizes as
well as longer run times should be investigated. Because real
faults are finite and the number of events observed is small, the
$\alpha$-dependence of $\bb$ seen in the \lr\ \BK\ model may accurately reflect
the behavior of some real faults. The qualitatively different behavior
observed for $\alpha \lesssim 1$ and $\alpha \gtrsim 1$ is consistent with recent results~\cite{Nagelgroup, clancy} for $R=1$.

For large $\alpha$ and large $R$, our results resemble those observed in
laboratory experiments on rocks. As the range
$R$ is increased and $\alpha$ is decreased, our results more closely
resemble the \lr\ CA models. This wide range of behavior indicates that the
physics of several models of earthquake
faults~\cite{BKModel67,CarlsonPRA89,RundleJSP91,OFCModelPRL92} can be
obtained from the generalized \BK\ model with the appropriate choice of $R$
and $\alpha$. Our results can be interpreted as suggesting that real earthquake faults and laboratory rocks
can have different statistical distributions of events and different physical
characteristics due to the details of the friction force as well as range of
the stress transfer. This dependence means that we must develop ways of
determining the friction force with considerably more accuracy if we want to
understand the relation between physical processes and observed earthquake
phenomena. Our waiting-time results also shows 
that other physical features might need to be added to 
the generalized \BK\ model considered in this paper such as memory effects, heterogeneities, and a more realistic rate-dependent friction force~\cite {ScholzBook02,pnas,ratestate}.

\begin{acknowledgements}

This work has been supported in part by DOE grants DE-FG02-95ER14498 (WK and
JX) and DE-FG02-04ER15568 (JBR). The simulations were done at Clark University with the partial 
support of NSF grant DBI-0320875.
\end{acknowledgements}


\begin{thebibliography}{99}
\bibitem{fixedloadingplate}Note that the loading plate is fixed and the substrate is moved to the left. The direction of the net force from the loading plate and the neighboring blocks is to the right. Hence, a block will begin to move 
to the right, the direction opposite to the moving substrate. The asymmetry 
of the friction force causes it to cancel the net force due to the loading plate and neighboring blocks so that a block will not be accelerated to the left. Hence, the friction force prohibits slip (relative to the substrate) in the same direction as the motion of the substrate.
\bibitem{zerovelocitylimit}Note that we do not load the system while the blocks are slipping.
\bibitem{creep}The results of Refs.~\onlinecite{Nagelgroup} and \onlinecite{clancy} suggest that a transition between creep and
stick-slip behavior might occur for $R=1$ and $\alpha=1$. Because the long-range \BK\ model is ergodic for small $\alpha$ and is not ergodic for large $\alpha$, in contrast to the $R=1$ model which is ergodic for all $\alpha$ studied, the mechanisms behind the different behavior near $\alpha = 1$ for $R=1$ and large $R$ might not be the same.
\bibitem{mori3}T. Mori and H. Kawamura, arXiv:0802.1305, have recently simulated the two-dimensional \BK\ model with a long-range interaction between blocks that decreases with the distance $r$ as $1/r^3$. Their qualitative results are consistent with ours.
\end{thebibliography}
\end{document}